\documentclass[useAMS,usenatbib,usegraphicx]{mn2e}

\usepackage{url}
\usepackage[varg]{txfonts}

\DeclareRobustCommand{\ion}[2]{%
  \textup{#1\,{\mdseries\textsc{#2}}}%
}

\renewcommand\vec[1]{\bmath{#1}}
\newcommand\mat[1]{\mathbf{#1}}

\newcommand{\veceta}{\vec{\eta}}

\newcommand{\vc}{v_{\mathrm{c}}}
\newcommand{\Rc}{R_{\mathrm{c}}}
\newcommand{\model}{\mathcal{M}}
\newcommand{\nTIC}{N_{\mathrm{TIC}}}
\newcommand{\nE}{N_{E}}
\newcommand{\nLz}{N_{L_{z}}}

\newcommand{\vphi}{\langle v_{\varphi} \rangle}
\newcommand{\evid}{\mathcal{E}}

\newcommand{\lamlen}{\lambda_{\mathrm{len}}}
\newcommand{\lamx}{\lambda^{\mathrm{dyn}}_{L}}
\newcommand{\lamy}{\lambda^{\mathrm{dyn}}_{E}}
\newcommand{\dynlen}{\textsc{cauldron}}
\newcommand{\galaxy}{SDSS\,J2321$-$097}
\newcommand{\Lz}{L_{z}}

\newcommand{\mt}{\tilde{m}}
\newcommand{\talp}{\alpha_{0}}
\newcommand{\PA}{\vartheta_{\mathrm{PA}}}
\newcommand{\Nrealiz}{{128}}

\newcommand{\qiso}{q_{\mathrm{iso, 2D}}}
\newcommand{\qlight}{q_{\mathrm{\star, 3D}}}
\newcommand{\Reff}{R_{\mathrm{eff}}}
\newcommand{\Leff}{L_{\mathrm{eff}}}
\newcommand{\Meff}{M_{\mathrm{eff}}}
\newcommand{\REin}{R_{\mathrm{Einst}}}
\newcommand{\MEin}{M_{\mathrm{Einst}}}

\title[Two-dimensional kinematics of SLACS lenses --
I.]{Two-dimensional kinematics of SLACS lenses -- I.~Phase-space
  analysis of the early-type galaxy SDSS\,J2321$-$097 at $\bmath{z
    \approx 0.1}$\thanks{Based on observations made with ESO
    Telescopes at the La Silla or Paranal Observatories under
    programme ID 075.B-0226 and 177.B-0682 and on observations made
    with the NASA/ESA \textit{Hubble Space Telescope}, obtained at the
    Space Telescope Science Institute, which is operated by the
    Association of Universities for Research in Astronomy, Inc., under
    NASA contract NAS 5-26555.}}

\author[O. Czoske et al.]{%
  Oliver Czoske$^{1}$\thanks{E-mail: czoske@astro.rug.nl},
  Matteo Barnab\`e$^{1}$,
  L\'eon V. E. Koopmans$^{1}$,
  Tommaso Treu$^{2}$ 
  \newauthor
  and Adam S. Bolton$^{3}$\\
  $^{1}$Kapteyn Astronomical Institute, PO Box 800, 9700\,AV
  Groningen, the Netherlands\\
  $^{2}$Department of Physics, University of California, Santa
  Barbara, CA 93101, USA\\
  $^{3}$Institute for Astronomy, University of Hawaii, 2680 Woodlawn
  Drive, Honolulu, HI 96822-1897, USA}

\begin{document}

\date{Accepted 2007 November 27. Received 2007 November 27; in
  original form 2007 November 6}

\pagerange{\pageref{firstpage}--\pageref{lastpage}} \pubyear{2007}

\maketitle

\label{firstpage}

\begin{abstract}
  We present the first results of a combined VLT VIMOS integral-field
  unit and \textit{Hubble Space Telescope} (\textit{HST})/ACS study of
  the early-type lens galaxy SDSS\,J2321$-$097 at $z=0.0819$,
  extending kinematic studies to a look-back time of 1\,Gyr.  This
  system, discovered in the Sloan Lens ACS Survey, has been observed
  as part of a VLT Large Programme with the goal of obtaining
  two-dimensional stellar kinematics of 17 early-type galaxies to $z
  \approx 0.35$ and Keck spectroscopy of an additional dozen lens
  systems.  Bayesian modelling of both the surface brightness
  distribution of the lensed source and the two-dimensional
  measurements of velocity and velocity dispersion has allowed us,
  under the only assumptions of axisymmetry and a two-integral stellar
  distribution function (DF) for the lens galaxy, to dissect this
  galaxy in three dimensions and break the classical mass--anisotropy,
  mass-sheet and inclination--oblateness degeneracies. Our main
  results are that the galaxy (i) has a total density profile well
  described by a single power-law $\rho \propto r^{-\gamma'}$ with
  $\gamma' = 2.06^{+0.03}_{-0.06}$; (ii) is a very slow rotator
  (specific stellar angular momentum parameter $\lambda_{\mathrm{R}} =
  0.075$); (iii) shows only mild anisotropy ($\delta \approx 0.15$);
  and (iv) has a dark-matter contribution of $\sim 30$~per cent inside
  the effective radius. Our first results from this large combined
  imaging and spectroscopic effort with the VLT, Keck and \textit{HST}
  show that the structure of massive early-type galaxies beyond the
  local Universe can now be studied in great detail using the
  combination of stellar kinematics and gravitational
  lensing. Extending these studies to look-back times where
  evolutionary effects become measurable holds great promise for the
  understanding of formation and evolution of early-type galaxies.
\end{abstract}

\begin{keywords}
  gravitational lensing ---
  techniques: spectroscopic ---
  galaxies: elliptical and lenticular, cD --- 
  galaxies: kinematics and dynamics --- 
  galaxies: structure.
\end{keywords}

\section{Introduction}
\label{sec:Introduction}

Within the hierarchical galaxy formation scenario, early-type galaxies
are assumed to be the end-products of major mergers with additional
accretion of smaller galaxies (e.g.~\citealt{Burkert-Naab2004} for a
review).  As such the study of their structure, formation and
evolution is an essential step in understanding galaxy formation and
the standard $\Lambda$ cold dark matter ($\Lambda$CDM) paradigm, in
which the hierarchical galaxy formation model has its
foundations. Early-type galaxies are observed to follow tight scaling
relations between their stellar population, dark matter, kinematic and
black hole properties, the origins of which are still not well
understood \citep[e.g.][]{Djorgovski-Davis1987, Dressler1987,
  Magorrian1998, Bolton2007}.

Considerable observational progress has been made in the last decades
in our understanding of the relative contributions of baryonic (mostly
stellar), dark matter and black hole constituents of early-type
galaxies through stellar dynamical tracers and X-ray studies (e.g.\
\citealt{Fabbiano1989}; \citealt{Mould1990}; \citealt*{Saglia1992};
\citealt{Bertin1994}; \citealt*{Franx1994}; \citealt{Carollo1995};
\citealt{Arnaboldi1996}; \citealt{Rix1997}; \citealt{Matsushita1998};
\citealt{Loewenstein-White1999}; \citealt{Gerhard2001};
\citealt{Seljak2002}; \citealt*{Borriello2003};
\citealt{Romanowsky2003}).  More recently, the SAURON collaboration
\citep{deZeeuw2002} has extensively studied a large and uniform sample
of early-type galaxies in the local Universe (i.e.\ $z \la 0.01$),
combining SAURON integral-field spectroscopy (IFS) on the William
Herschel Telescope with high-resolution \textit{Hubble Space
  Telescope} (\textit{HST}) imaging to obtain a three-dimensional
picture of these galaxies in terms of their mass and kinematic
structure \citep{Emsellem2004, Emsellem2007, Cappellari2006,
  Cappellari2007}.

\begin{figure*}
  \centering
  \resizebox{0.375\hsize}{!}{\includegraphics{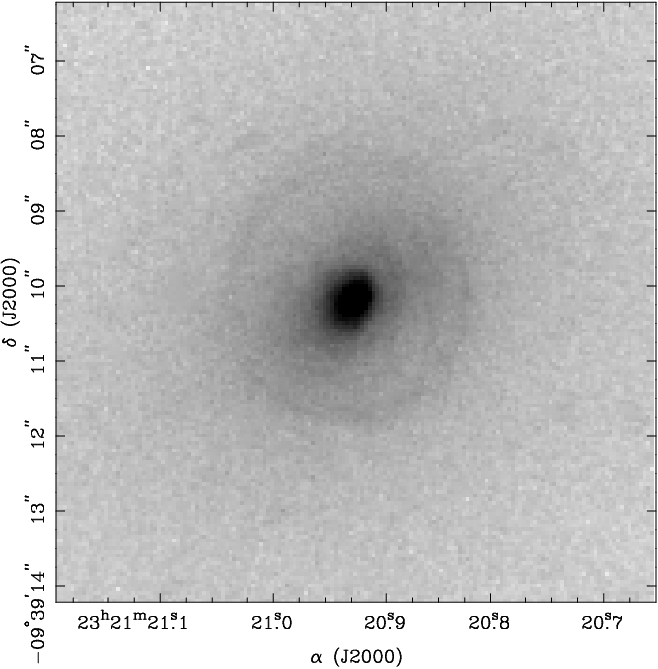}}
  \hspace{0.025\hsize}
  \resizebox{0.375\hsize}{!}{\includegraphics{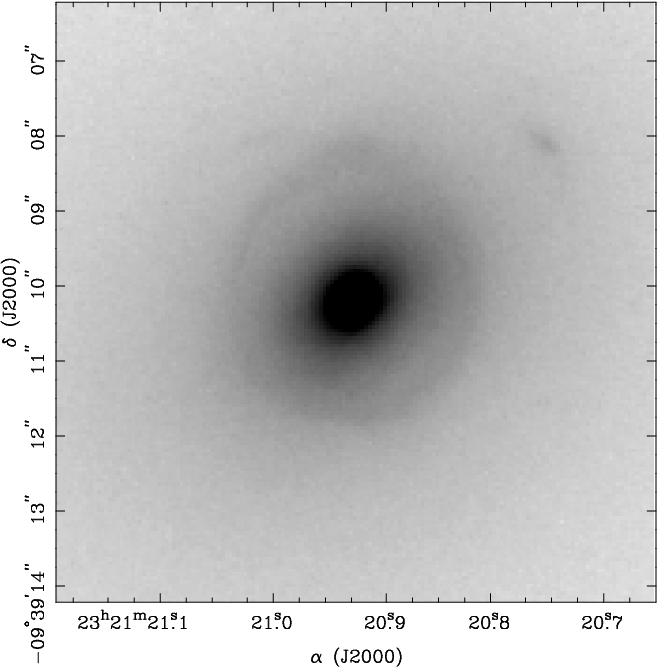}}
  \caption{\textit{HST}/ACS images of \galaxy, taken through the
    \textit{F435W} (left-hand panel) and \textit{F814W} (right-hand
    panel) filters. The images are single exposures with an
    integration time of $420\,\mathrm{s}$; the pixel scale is
    $0.05\,\mathrm{arcsec}$. A full description of these observations
    and the data-reduction process is given in \citet{Bolton2006}.}
  \label{fig:HST_images}
\end{figure*}

Despite making a great leap forward, the complex modelling of the
SAURON systems and the unknown dark matter content and total mass of
these galaxies still require some assumptions to be made about their
gravitational potential [e.g.\ constant stellar mass-to-light ratio
($M/L$) from stellar population models] and inclination. These issues
can be partly overcome in discy systems where the inclination can be
determined from the photometry or possibly from dust lanes.  However,
systems with lower ellipticity can show stronger degeneracies
\citep[e.g.][]{Cappellari2007}. Also, near the effective radius dark
matter becomes a major contributor to the stellar kinematics
\citep{Gerhard2001, Treu-Koopmans2004} and it is not yet clear what
impact its neglect has, even though it is estimated to be not too
severe \citep{Cappellari2006}.

At higher redshifts, the extraction of detailed kinematic information
from early-type galaxies becomes increasingly more difficult because
of the decrease in apparent size and cosmological surface brightness
dimming.  However, galaxies at higher redshifts are much more likely
to act as gravitational lenses on background galaxies
\citep*[e.g.][]{Turner1984}.  The additional information obtained from
gravitational lensing on, for example, the enclosed mass and the
density profile has been shown to break some of the classical
degeneracies in stellar kinematics and lensing, namely the
mass--anisotropy and the mass-sheet degeneracies, respectively
\citep[e.g.][hereafter BK07]{Treu-Koopmans2002a, Koopmans2003,
  Koopmans-Treu2003, Treu-Koopmans2004, Koopmans2006,
  Barnabe-Koopmans2007}.

In order to use gravitational lensing and stellar dynamics in a
systematic way, the Lenses Structure \& Dynamics (LSD) Survey was
started in 2001, combining \textit{HST} imaging data on lens systems
with stellar kinematics obtained with Keck \citep{Koopmans-Treu2002,
  Koopmans-Treu2003, Treu-Koopmans2002a, Treu-Koopmans2002b,
  Treu-Koopmans2003, Treu-Koopmans2004}. This project and its results,
predominantly at $z\approx 0.5\!\!-\!\!1.0$, have shown the validity
of this methodology and its effectiveness at higher redshifts. These
high-redshift systems, however, are relatively rare and hard to follow
up, which has limited studies to the use of luminosity-weighted
stellar velocity dispersions
\citep[e.g.][]{Koopmans-Treu2002,Treu-Koopmans2002b} or long-slit
spectroscopy in a few large apertures along the major axes
\citep[e.g.][]{Koopmans-Treu2003, Treu-Koopmans2004}.

The Sloan Lens ACS Survey \citep[SLACS,][]{Bolton2005,Bolton2006} was
begun to extend the sample of lens galaxies suitable for joint lensing
and dynamical studies. SLACS lens candidates were selected from the
SDSS Luminous Red Galaxy sample \citep{Eisenstein2001} and a quiescent
subsample [defined by $\mathit{EW}(\mathrm{H}\alpha)<1.5\,$\AA] of the
MAIN SDSS galaxy sample \citep{Strauss2002} by an inspection of their
SDSS fibre spectroscopy \citep{Bolton2004}. Galaxies whose spectra
presented emission lines at a higher redshift than the redshift of the
galaxy itself were observed in an \textit{HST} snapshot programme to
check whether the source of the emission lines was a gravitationally
lensed background galaxy. To date, SLACS has confirmed around 80~lens
systems (Bolton et al., in preparation). In contrast to most earlier
lens surveys which drew their candidate systems from catalogues of
potential sources \citep[e.g.~quasi-stellar objects or radio sources
as in the CLASS survey,][]{Browne2003}, SLACS is a
\emph{lens}-selected survey and as such preferentially finds systems
with bright lens galaxies but comparatively faint sources. It is
therefore an ideal starting point for detailed investigations into the
properties of the lens galaxies.

In this paper we introduce a follow-up project to SLACS which aims at
combining the lensing information with detailed two-dimensional
kinematic information obtained with the VIMOS integral-field unit
(IFU) mounted on the VLT.  The full sample comprises 17~systems. In
this paper we demonstrate the available data and the analysis
methodology on one system, {\galaxy}. The lens in this system is an
early-type galaxy at $z=0.0819$. The SDSS spectrum shows additional
[\ion{O}{ii}] and H$\beta$ emission from a gravitationally
lensed blue galaxy at $z_{\mathrm{s}} = 0.5342$ \citep{Bolton2006}.
In \textit{HST} imaging (Fig.~\ref{fig:HST_images}) the lensed source
is shown to form an almost complete Einstein ring of radius
$1.68\,\mathrm{arcsec}$, corresponding to an Einstein radius of
$\REin=2.6\,\mathrm{kpc}$ at the lens redshift. The projected mass
enclosed within the Einstein radius is $M(<\REin) =
1.3\times10^{11}\,M_{\sun}$.  Details on this system are listed in
Table~\ref{tab:basic_data}.

In Section~\ref{sec:Observations}, we describe the VIMOS and
\textit{HST} observations that form the basis of this analysis. The
IFU data are presented in Section~\ref{sec:data_analysis}, with a
description of the data reduction in Section~\ref{ssec:data_reduction}
and of the extraction of two-dimensional kinematic maps in
Section~\ref{sec:kinematic_analysis}.  In Section~\ref{sec:inferences}
we determine several commonly used parameters from the kinematic data
alone, in a way that is directly comparable to earlier studies, for
example from the SAURON collaboration. We apply the joint lensing and
dynamics analysis developed by BK07 in Section~\ref{sec:analysis}. In
Section~\ref{sec:degeneracies} we present a physical argument that
explains why gravitational lensing and stellar dynamics can break the
degeneracy between oblateness and inclination. We discuss our results
and conclude in Section~\ref{sec:discussion}.
We use a concordance $\Lambda$CDM model throughout this paper,
described by $\Omega_{\mathrm{M}}=0.3$, $\Omega_{\Lambda} = 0.7$ and
$H_{0} = 100\,h\,\mathrm{km\,s^{-1}\,Mpc^{-1}}$ with $h=0.7$ unless
stated otherwise. At the redshift of \galaxy, $1\,\mathrm{arcsec}$
corresponds to $1.1\,h^{-1}\,\mathrm{kpc} = 1.5\,\mathrm{kpc}$.

\vfill

\begin{table}
  \centering
  \begin{minipage}{0.85\hsize}
  \caption{Basic data for \galaxy. The data are taken from
    \citet{Treu2006err} and \citet{Koopmans2006}.}
  \begin{tabular}{@{}l@{\hspace{12em}}r@{$\;$}l}
    \hline
    $\alpha_{\mathrm{J2000}}$ &
    $23^{\mathrm{h}}21^{\mathrm{m}}20\fs93$ & \\
    $\delta_{\mathrm{J2000}}$ & $-09\degr39\arcmin10\farcs2$ & \\
    $z_{\mathrm{l}}$ & $0.0819$ & \\
    $z_{\mathrm{s}}$ & $0.5342$ & \\
    $\sigma_{\mathrm{ap}}$ & $(236\pm7)$&$\mathrm{km\,s^{-1}}$ \\
    $\sigma_{\mathrm{c}}$ & $(245\pm7)$&$\mathrm{km\,s^{-1}}$ \\
    $\REin$ & $1.68\,\mathrm{arcsec} = 2.59$ & $\mathrm{kpc}$  \\
    $R_{\mathrm{eff},B}$ & $(8.47\pm0.11)$&$\mathrm{kpc}$ \\
    $M_{B}$ & $-21.72\pm0.05$ & \\
    $R_{\mathrm{eff},V}$ & $(7.93\pm0.07)$&$\mathrm{kpc}$ \\
    $M_{V}$ & $-22.59\pm0.05$ & \\
    $\qiso$   & $0.77$ & \\
    $\vartheta_{\mathrm{PA},\star}$ & $126\fdg5$ & \\
    \hline
  \end{tabular}
  \textit{Notes:} $z_{\mathrm{l}}$ and $z_{\mathrm{s}}$ are the
  redshifts of \galaxy\ and the gravitationally lensed background
  galaxy, respectively. $\sigma_{\mathrm{ap}}$ is the velocity
  dispersion measured from the 3-arcsec-diameter SDSS fibre,
  $\sigma_{\mathrm{c}}$ is the derived central velocity dispersion
  \citep{Treu2006}. $\REin$ is the Einstein radius. $R_{\mathrm{eff}}$
  and $M$ are the effective radii and absolute magnitudes determined
  by fitting de Vaucouleurs profiles to the $B$- and $V$-band ACS
  images \citep{Treu2006}. $\qiso$ and $\vartheta_{\mathrm{PA},\star}$
  are isophotal axial ratio and position angle of the major axis,
  respectively \citep{Koopmans2006}.
  \end{minipage}
  \label{tab:basic_data}
\end{table}

\section{Observations}
\label{sec:Observations}

In this section, we briefly discuss the VLT VIMOS/IFU spectroscopic
and the \textit{HST} imaging observations, which form the basic data
sets for the subsequent modelling.

\subsection{Integral-field spectroscopy}
\label{ssec:instrument}

We have obtained integral-field spectroscopic observations of
17~systems using the IFU of VIMOS \citep{LeFevre2001} mounted on the
VLT UT3 (Melipal). The observations were conducted in two parts: three
systems were observed in a pilot programme (ESO programme 075.B-0226; PI:
Koopmans) and the remaining 14 systems in a large programme (177.B-0682;
PI: Koopmans) with a slightly different observational
setup. Observations were carried out in service mode throughout to
ensure uniform observing conditions. In this paper we report in
detail on the observations of the pilot programme and the analysis of
one system, \galaxy. The observations of the large programme and the
analysis of the full sample will be presented in forthcoming papers in
this series.

The IFU head of VIMOS consists of a square array of $80\times80$
microlenses with a sampling of $0.67\,\mathrm{arcsec}$ per spatial
element (`spaxel') at the Nasmyth focus of the telescope. In
high-resolution mode only the central 1600 lenses are used, resulting
in a field of view of $27\times27\,\mathrm{arcsec^{2}}$. The lenses
are coupled to optical fibres which transmit the light to pseudo-slit
masks, each of which contains 400~fibres and corresponds to a quadrant
of the original field. The seeing limit for the observations was set
to $0.8\,\mathrm{arcsec}$, so that each fibre of the IFU gives a
spectrum that is essentially independent of its neighbours.

For the pilot programme, we employed the HR-Blue grism which covers a
wavelength range from $4000$ to $6200$\,\AA\ (the exact wavelength
range varies slightly from quadrant to quadrant).  The spectral
resolution of the HR-Blue grism is $\lambda/\Delta \lambda = 2550$,
where $\Delta \lambda$ is the full width at half-maximum (FWHM) of the
instrument profile. The mean dispersion on the CCD is $0.5$\,\AA\ per
pixel. Three exposures of $555\,\mathrm{s}$ integration time were
obtained per observing block (a one-hour sequence of science and
calibration exposures); the telescope was shifted by about
$3.5\,\mathrm{arcsec}$ between the exposures to fill in on dead
fibres.  The number of observing blocks spent on each lens system was
adjusted so as to reach approximately equal signal-to-noise ratios
($S/N$) across the sample. For the system described in this paper,
SDSS\,J2321$-$097, five observing blocks were carried out for a total
integration time of $8325\,\mathrm{s}$.  For the large programme, we
switched to the HR-Orange grism which covers a wavelength range from
$5250$ to $7400$\,\AA\ at a similar spectral resolution of
$\lambda/\Delta \lambda = 2650$ and a dispersion of $0.6$\,\AA\ per
pixel.  The spectral resolution translates to a rest-frame velocity
resolution $\Delta v$ varying between $107$ and
$85\,\mathrm{km\,s^{-1}}$, improving as the redshift of our sample
lens galaxies increases from $0.08$ to $0.35$.

\subsection{\textit{HST} imaging} 
\label{ssec:HST_imaging}

High-resolution imaging with \textit{HST}/ACS or \textit{HST}/WFPC2
has been or is being obtained for all confirmed SLACS lenses
\citep{Bolton2006,Gavazzi2007}. The data used here come from the ACS
snapshot survey described by \citet{Bolton2005} and consist of single
$420\,\mathrm{s}$ exposures through the \textit{F435W} and
\textit{F814W} filters, respectively. Fig.~\ref{fig:HST_images} shows
the images of \galaxy.  The lens modelling requires as input a clean
estimate of the brightness distribution of the lensed images, so the
smooth brightness distribution of the lens galaxy has to be subtracted
from the images.  \citet{Bolton2006} tried a variety of commonly used
modelling techniques (e.g.~fitting of Sersic profiles), but settled on
the more flexible B-spline technique. The galaxy-subtracted image in
the \textit{F814W} filter is shown in the top right-hand panel of
Fig.~\ref{fig:LENcomp}.

\section{IFU data analysis}
\label{sec:data_analysis}

\subsection{Data reduction}
\label{ssec:data_reduction}

\begin{figure*}
  \centering
  \resizebox{0.375\hsize}{!}{\includegraphics{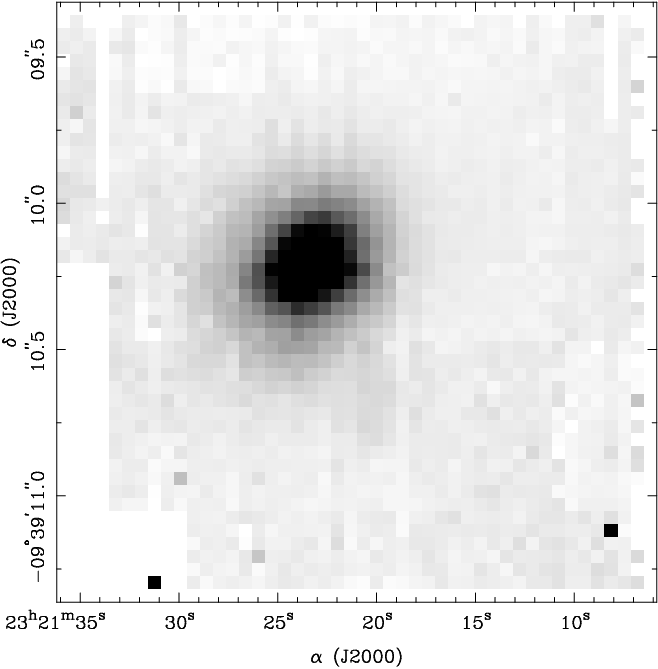}}
  \hspace{0.025\hsize}
  \resizebox{0.375\hsize}{!}{\includegraphics{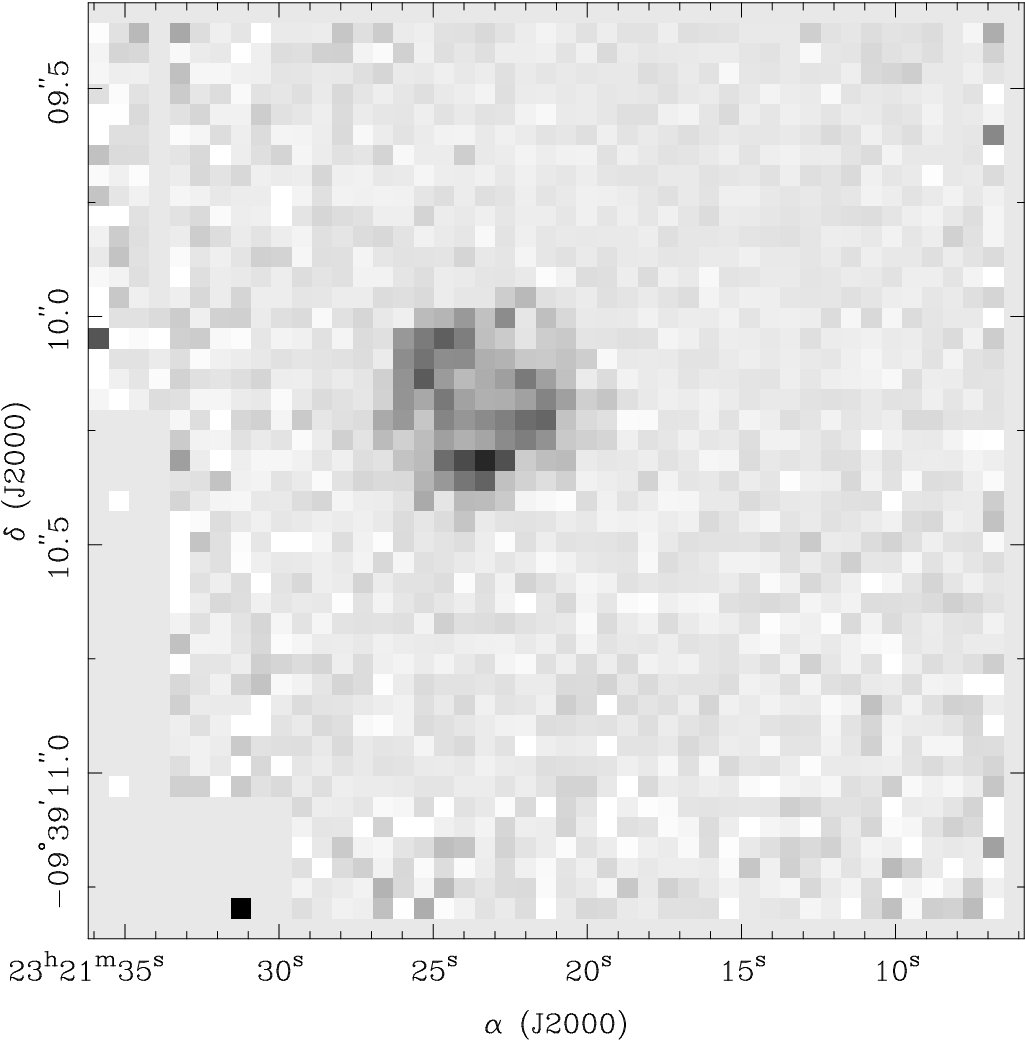}}
  \caption{Left-hand panel: VLT VIMOS/IFU image reconstruction of
    SDSS\,J2321$-$097, obtained by integrating along the individual
    fibre spectra from 4400 to 6100\,\AA. Right-hand panel: map of the
    strength of the [\ion{O}{ii}] doublet from the lensed source.}
  \label{fig:image}
\end{figure*}

The VLT VIMOS/IFU data were reduced using the \textsc{vipgi} package
which was developed within the framework of the VIRMOS consortium and
the VVDS project.
\textsc{vipgi}\footnote{http://cosmos.iasf-milano.inaf.it/pandora/vipgi.html}
has been described in detail by \citet{Scodeggio2005} and
\citet{Zanichelli2005}; here, we restrict ourselves to a brief summary
of the main reduction steps performed by \textsc{vipgi}.

The four quadrants of the IFU are treated independently up to the
combination of all exposures into the final data cube.  For each night
on which observations were carried out, five bias exposures were
obtained which are median-combined and subtracted from the data.  The
observed halogen-lamp illuminated screen flats are used to locate the
traces of the individual spectra on the CCD, starting from
interactively adjusted first guesses of the layout of the fibres. This
includes a shift correction and an identification of `bad' fibres. The
position of the peak of each spectrum in the spatial direction is then
modelled as a quadratic polynomial function of the position in the
dispersion direction. \textsc{vipgi} does not perform a CCD
pixel-to-pixel flat-field correction \citep{Scodeggio2005}.

The wavelength calibration is subsequently done using helium and neon
lamp observations taken immediately after the science exposures. With
about 20~lines spread between $\sim4000$ and $\sim 6200$\,\AA, the rms
deviations from the polynomial fit are roughly Gaussian-distributed
around a mean of $\sim 0.075$\,\AA\ and a dispersion of $\sim
0.02$\,\AA. The velocity error introduced by uncertainties in the
wavelength calibration is thus of the order of
$5\,\mathrm{km\,s^{-1}}$, significantly smaller than the spectral
resolution and negligible in our analysis.

\begin{figure*}
  \centering
  \resizebox{0.75\hsize}{!}{\includegraphics{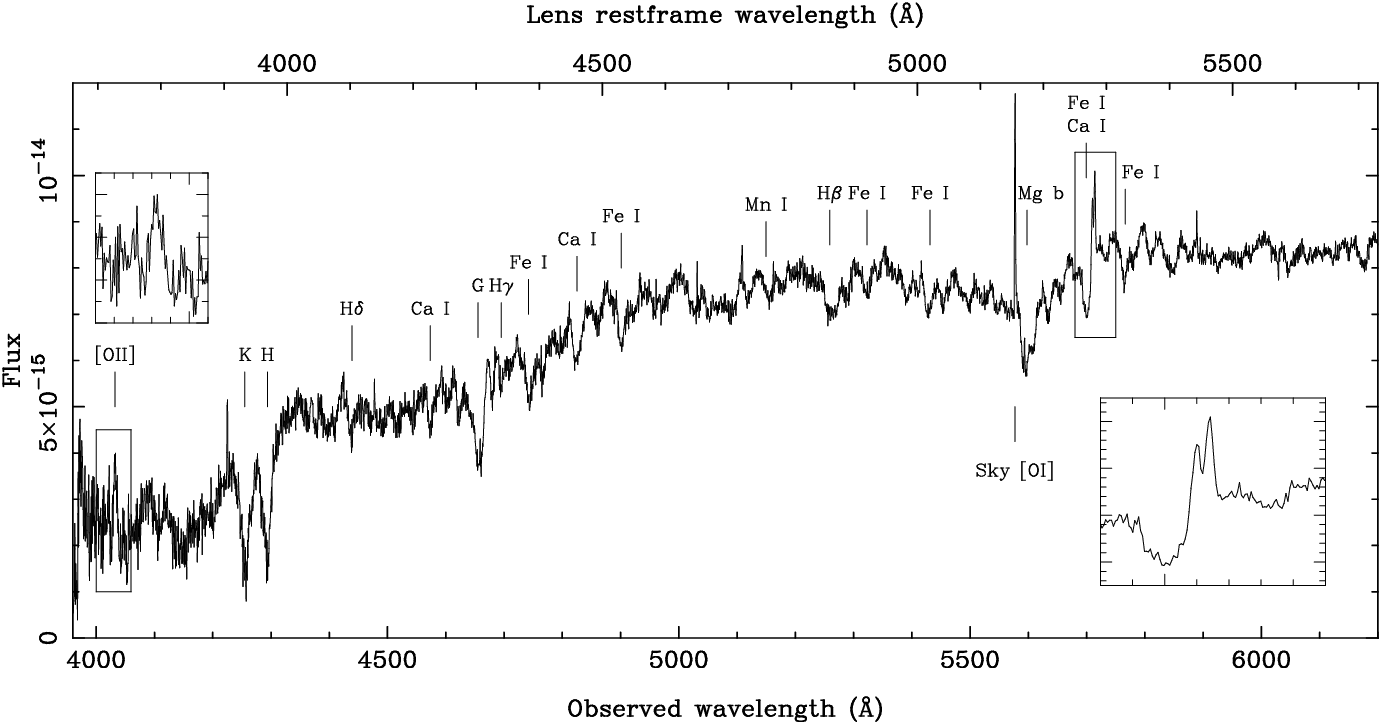}}
  \caption{Spectrum of {\galaxy} integrated from the VIMOS/IFU data
    cube over a circular aperture of diameter
    $3\,\mathrm{arcsec}$. The inset on the right-hand side shows a
    zoom on the [\ion{O}{ii}] doublet of the lensed background galaxy
    at $z_{\mathrm{s}}=0.5342$, next to the FeCa $\lambda 5268$ line
    of the lens. The inset on the left-hand side shows a zoom on the
    weak but clearly detected [\ion{O}{ii}] line of the lens
    galaxy. Several other spectral features of the lens are marked.}
  \label{fig:global_spectrum}
\end{figure*}

The individual spectra are extracted using the optimal extraction
scheme of \citet{Horne1986}, that is, by averaging the flux across the
spectral trace of the fibre, weighted by a fit to its spatial
profile. In order to correct for the transmissivity variations from
fibre to fibre (`spaxel-to-spaxel flat-fielding'), the flux under a
sky emission line is measured and scaled relative to a reference
fibre. We use [\ion{O}{i}] at $\lambda=5577$\,\AA\ for this purpose.

Spectrophotometric standard stars were observed during most
nights. Since the stellar light only covers a small fraction of the
fibres in the IFU field whereas the calibration is applied to all
fibres, the spectrophotometric calibration only accurately corrects the
relative wavelength dependence of the sensitivity function (assuming
that this is the same for all fibres), while the absolute flux levels
are determined with a much larger uncertainty due to residuals from
the correction of the varying fibre transmissivities.

The VIMOS IFU does not have dedicated sky fibres, but the
field of view is large enough for a sufficient number of fibres to
essentially measure pure sky. \textsc{vipgi} identifies these fibres
in a statistical way by integrating the flux in each fibre over a
large wavelength range and selecting as sky fibres those with a flux
within $\pm 0.1$~per cent of the mode of the resulting distribution.

Finally, the four quadrants of all the individual exposures are
combined into a single data `cube' after correction for the dithering
between exposures. The final data are in row-stacked form, where each
row corresponds to the spectrum from one fibre. A fits table extension
gives the translation from row number to the spatial coordinate of the
fibre.

The left-hand panel of Fig.~\ref{fig:image} shows an image of \galaxy,
reconstructed from the reduced data cube using the program
\textsc{sadio}\footnote{http://cosmos.iasf-milano.inaf.it/pandora/sadio.html}
by integrating along the wavelength direction from $4400$ to
$6100\,$\AA. The galaxy surface brightness distribution agrees well
with that seen in the \textit{HST}/ACS \textit{F435W} and
\textit{F814W} snapshot images, although it is not as precise due to
the complexities involved in the spectral calibration, in particular
of the relative fibre transmissivity. In the modelling of the
lens galaxy brightness distribution, we therefore use the \textit{HST}
images. The imperfect absolute flux calibration of the IFU data has no
effect on the measurements of the stellar kinematics which do not
depend on the overall scaling of each fibre spectrum.

Following \citet{Bolton-Burles2007}, we can use the IFU data to obtain
an image of the lensed source by isolating its redshifted
[\ion{O}{ii}] emission line at $5710\,$\AA.  The [\ion{O}{ii}] doublet
is modelled as the sum of two Gaussians with separation $4.5$\,\AA\
(taking into account the redshift of the source) and relative strength
3:2. In regions with significant light from the lens galaxy the
best-fitting kinematic model (Section~\ref{sec:kinematic_analysis})
was subtracted prior to the fit; elsewhere a polynomial fit was
subtracted to remove a remaining continuum level. The map of the line
strength of the [\ion{O}{ii}] line obtained in this manner is
presented in the right-hand panel panel of Fig.~\ref{fig:image}. The
map clearly shows the ring-like structure of the lensed galaxy
(cf.~the higher resolution galaxy-subtracted ACS image in the top
right-hand panel of Fig.~\ref{fig:LENcomp}) and demonstrates that the
[\ion{O}{ii}] line used for the inclusion of this system in the SLACS
sample indeed originates from a gravitationally lensed background
galaxy.

The spectral energy distribution of \galaxy\ is essentially that of a
typical early-type galaxy with strong metal lines, as seen in
Fig.~\ref{fig:global_spectrum} which shows the spectrum integrated
over a circular aperture of $3\,\mathrm{arcsec}$ diameter (i.e.~the
SDSS aperture). We note, however, that this spectrum shows weak but
significant [\ion{O}{ii}] emission ($\mathrm{EW}\sim-4\,$\AA) as well
as H$\delta$ absorption from the lens galaxy. This might indicate
residual star formation or active galactic nucleus activity at a low
level.

\subsection{Kinematic analysis}
\label{sec:kinematic_analysis}

We extract two-dimensional maps of systematic velocity $v(\vec{R})$
and line-of-sight velocity dispersion $\sigma(\vec{R})$ using a direct
pixel-fitting method implemented in the statistical language
\textsc{r}\footnote{http://www.r-project.org/} \citep{R}.  Since the
analysis is done on spectra from individual IFU fibres, the spectra
have relatively low signal-to-noise ratio, in particular in the
outskirts of the lens galaxy where the surface brightness drops
rapidly. We therefore restrict the model to a Gaussian line-of-sight
velocity distribution (LOSVD) and do not attempt to measure higher
order moments, such as the Gauss--Hermite coefficients $h_{3}$ and
$h_{4}$ commonly used in kinematic studies of low-redshift early-type
galaxies
\citep[e.g.][]{vanderMarel-Franx1993,Cappellari-Emsellem2004}.

The observed galaxy spectrum $d_{i}(\lambda)$ is modelled, for each
fibre $i$ separately, by the convolution of a high-resolution stellar
template spectrum $t(\ln \lambda)$ with a Gaussian kernel $g_{i}(\ln
\lambda)$ with dispersion $\sigma_{v,i}/c$:
\begin{equation}
  \label{eq:model}
  d_{i}(\lambda) = f_{i}(\lambda)\,[ t\otimes g_{i} ](\ln \lambda) 
  + n_{i}(\lambda)\;.
\end{equation}
We assume Gaussian noise $n(\lambda)$ on the spectrum.  Since the
accuracy of the spectrophotometric calibration is not perfect for both
our spectra and the chosen template spectra \citep[see below
and][]{Valdes2004} there remain small relative tilts between the
continua of the galaxy and template spectra.  We model this difference
by the linear correction function $f(\lambda) = a + b\lambda$ whose
parameters are determined by a linear fit nested within the non-linear
optimization for the kinematic parameters $v_{i}$ and $\sigma_{v,i}$.

The template spectrum is first resampled to a logarithmic wavelength
scale with step $\Delta \ln \lambda = \Delta v/c$, where we use a
velocity step of $\Delta v = 25\,\mathrm{km\,s^{-1}}$. The convolution
kernel is a Gaussian with dispersion
$\sigma_{\mathrm{G}}=\sigma_{v}/\Delta v$, extending to
$\pm8\sigma_{\mathrm{G}}$.  The galaxy spectrum is corrected for the
mean redshift of the lens galaxy, $z=0.0819$ in the case of \galaxy,
taken from the SDSS data base.  The model is then resampled on to the
wavelength grid of the galaxy spectrum, a linear wavelength scale
$\lambda_{\mathrm{obs}}/(1+z_{\mathrm{SDSS}})$ in our case. This
avoids further resampling of the data beyond what was done in the data
reduction. Similar methods typically resample both model and data to
the same logarithmic wavelength grid before comparing
\citep[e.g.][]{vanderMarel1994,Cappellari-Emsellem2004}.

\begin{figure}
  \centering
  \resizebox{0.9\hsize}{!}{\includegraphics{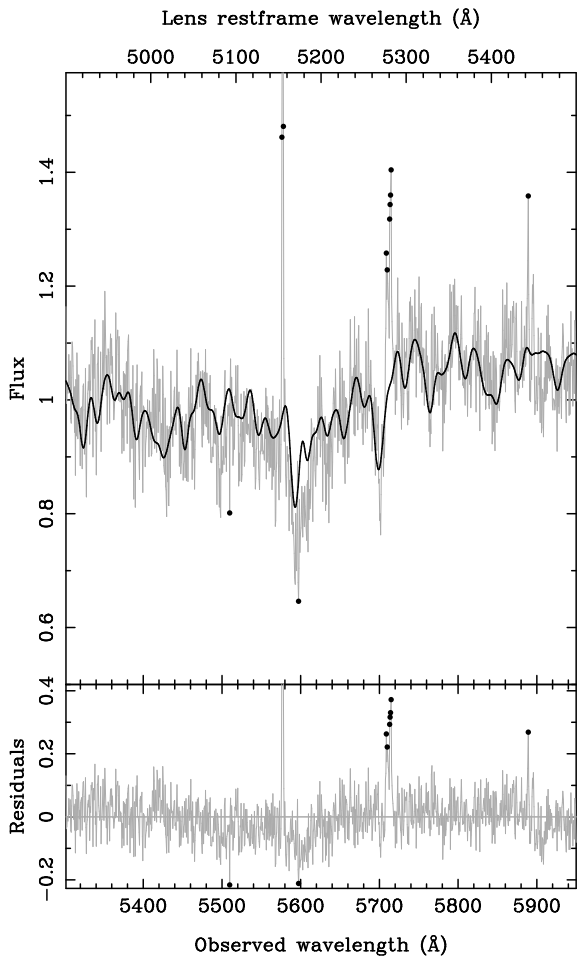}}
  \caption{Spectrum from a single fibre near the centre of {\galaxy},
    with the overlaid best-fitting model (top panel). The bottom panel
    shows the residuals of the fit. The [\ion{O}{ii}] doublet from the
    source at $5710\,$\AA\ ($z_{\mathrm{s}}= 0.5342$) and the residual
    of the [\ion{O}{i}] sky line at $5577\,$\AA\ were masked
    automatically (indicated by the points).}
  \label{fig:model:plot}
\end{figure}

Remaining cosmic rays, emission lines, and remnants of sky lines are
masked automatically by $3\sigma$-clipping of the residuals from a
preliminary fit without mask. Fig.~\ref{fig:model:plot} shows an
example of a fit to a spectrum from a fibre near the centre of
\galaxy.  Error analysis is performed through a Monte Carlo technique
by adding Gaussian noise to the best-fitting model at the noise level
given by the residuals between model and data, and using 1000
realizations to determine the spread of the best-fitting parameters.

The choice of a good template spectrum is important for accurate
kinematic measurements from galaxy spectra. For \galaxy, we started by
experimenting with a set of eight stellar template spectra obtained
with the Echelle Spectrograph and Imager (ESI) on the Keck~II
telescope \citep{Koopmans-Treu2002} applied to the global spectrum of
\galaxy.  These spectra have a velocity resolution of $\Delta
v_{\mathrm{ESI}} = 22.4\,\mathrm{km\,s^{-1}}$ (ESI
manual\footnote{http://www2.keck.hawaii.edu/inst/esi/echmode.html})
and cover a range of spectral types G0 to K4, luminosity class III.
Fitting to the integrated spectrum of \galaxy\ it was found that four
templates for stellar types between G7 and K0 gave consistent results
for the velocity dispersion within $\pm10\,\mathrm{km\,s^{-1}}$, which
also agreed with the measurement from the SDSS. Cooler template stars
(K2 and K4) resulted in a significantly higher velocity dispersion
while hotter templates (G0 and G5) gave lower values. On the basis of
the visual appearance of the fit and the shape of the galaxy spectrum
it was decided to use the K0 star HR\,19 as the template star for
\galaxy.

We subsequently exchanged the ESI spectrum of HR\,19 for the
corresponding spectrum from the Indo-US library of Coud\'e feed
stellar spectra\footnote{http://www.noao.edu/cflib/}
\citep{Valdes2004}. The Indo-US library contains spectra for 1273
stars covering a wide range of stellar parameters. The spectra have a
resolution of $\sim1\,$\AA\ FWHM and for the majority of stars cover a
wavelength range of $3460$ to $9464$\,\AA. While the resolution of the
Indo-US spectra is somewhat lower than that of the ESI spectra, this
disadvantage is offset by the large number of spectra available, which
will provide freedom of choice for the analysis of our full sample. We
have checked that the kinematic results obtained with the Indo-US and
the ESI templates are indeed consistent.

Fig.~\ref{fig:model:plot} shows an important limitation of any use of
observed stellar spectra as templates in kinematic analyses, namely
the differing abundance ratios between stars in the solar
neighbourhood and early-type galaxies \citep*{Barth2002}. While the
strengths of the \ion{Fe}{i}/\ion{Ca}{i}\,$\lambda 5268$ lines agree,
the Mg\,$b$\,$\lambda 5174$ line is significantly enhanced in the
galaxy spectrum. \citet{Barth2002} recommend masking the Mg\,$b$ line
before fitting. Doing this increases the velocity dispersion only by a
few per cent and does not significantly improve the fit outside the
Mg\,$b$ region.

\subsection{Results from the observed stellar kinematics}
\label{sec:inferences}

A fit of a kinematic model to the global spectrum
(Fig.~\ref{fig:global_spectrum}) following the procedure described in
Section~\ref{sec:kinematic_analysis} results in a velocity dispersion
of $\sigma_{\mathrm{tot}}= 240\pm8\,\mathrm{km\,s^{-1}}$, which is
consistent with the value $\sigma_{\mathrm{SDSS}} =
236\pm7\,\mathrm{km\,s^{-1}}$ listed in the SDSS data base
\citep{Bolton2006}.  Note that the error estimates are not directly
comparable because of differences in the estimation method. We believe
our Monte Carlo estimate to be more reliable, although it might still
underestimate the real error due to the assumption of Gaussian noise
in the spectrum \citep{DeBruyne2003}.

\begin{figure*}
  \centering
  \resizebox{!}{0.5\textheight}{\includegraphics{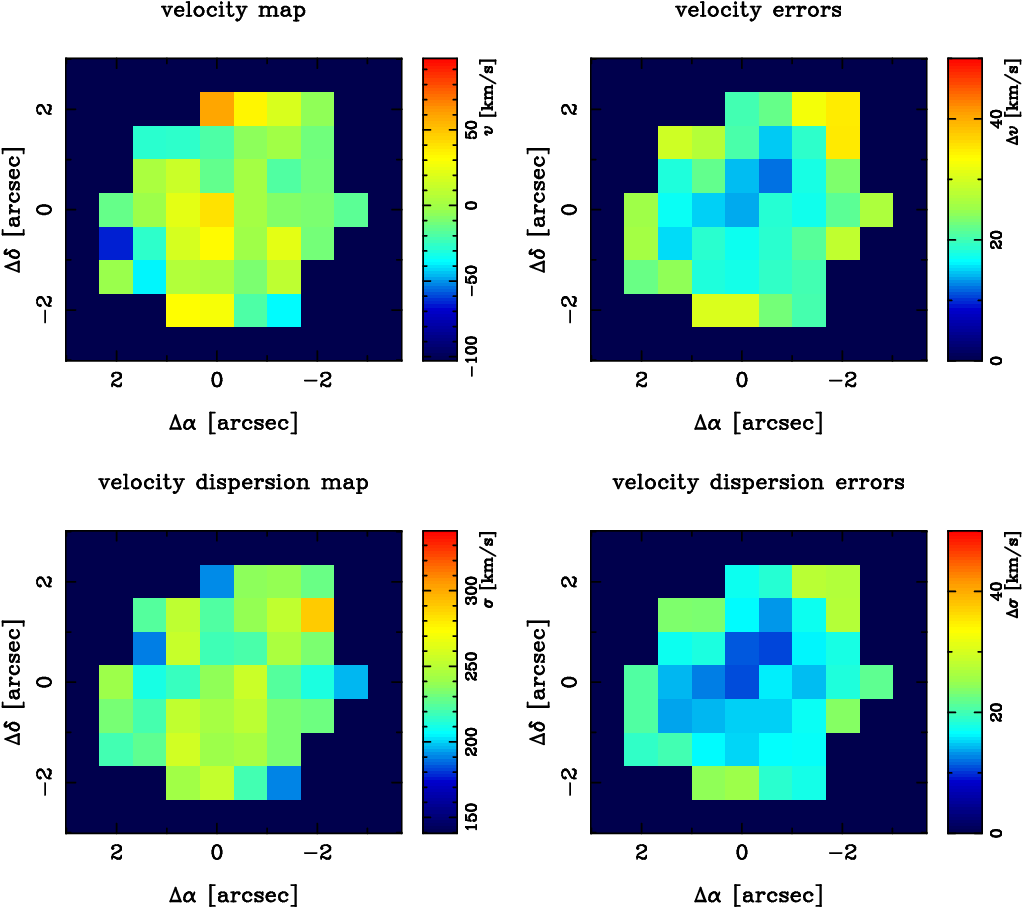}}
  \caption{Kinematic maps of \galaxy. The top left-hand panel shows
    systematic velocity relative to the mean redshift of the lens
    galaxy; the bottom left-hand panel shows the velocity
    dispersion. The panels on the right-hand side visualize the
    corresponding $1\sigma$ errors on the measurements. Spaxel size is
    $0.67\,\mathrm{arcsec}$ and only spectra with $S/N > 8$ were
    analysed.}
  \label{fig:kinematic_maps}
\end{figure*}

The two-dimensional maps of the velocity and velocity dispersion are
shown in Fig.~\ref{fig:kinematic_maps}. Only fibres in which the
spectrum has an $S/N$ per pixel in the rest-frame wavelength range
from 5350\,\AA\ to 5450\,\AA\ (a flat part of the spectrum) of $S/N>8$
are used in these maps and the subsequent analysis. Spectra with lower
$S/N$ were found not to yield reliable kinematic fits.

Neither the velocity nor the velocity dispersion map shows significant
structure. The velocity map might hint at a pattern of slight
rotation, although within the noise the map is consistent with no
rotation at all.

In order to quantify the amount of rotation in a galaxy from
two-dimensional maps of line-of-sight stellar velocity $v$ and
velocity dispersion $\sigma_{v}$, \citet{Emsellem2007} defined a
stellar angular momentum parameter $\lambda_{\mathrm{R}}$ by
\begin{equation}
  \label{eq:lambda_R:definition}
  \lambda_{\mathrm{R}} = \frac{\langle R\,|v|\rangle}
  {\langle R\sqrt{v^{2} + \sigma_{v}^{2}}\rangle}\;,
\end{equation}
where $R$ is the projected distance from the galaxy centre. We find
$\lambda_{\mathrm{R}} = 0.075$, which in combination with the absolute
magnitude $M_{B}=-21.72$ places \galaxy\ among the slowly rotating
galaxies at the high-luminosity end of the SAURON sample
\citep[cf.~fig.~7 of][]{Emsellem2007}.

Following the prescription of \citet{Binney2005} and
\citet{Cappellari2007}, we find a low value for the ratio between
systematic velocity and velocity dispersion of $(\langle
v^{2}\rangle/\langle \sigma_{v}^{2}\rangle)^{1/2} = 0.1$, which in
combination with the observed ellipticity of
$\epsilon_{\mathrm{iso,2D}} = 1-\qiso = 0.23$ from isophote fitting
places \galaxy\ well below the curve for an isotropic rotating oblate
galaxy in an $\epsilon\!\!-\!\!(v/\sigma_{v})$ diagram.  From the
tensor virial theorem the global anisotropy parameter
\citep{Binney2005} is found to be $\delta \sim 0.15$, where no
correction for inclination has been made yet. In
Section\,\ref{sec:analysis}, we calculate the same quantity from the
reconstructed stellar DF and find effectively the same result. Again,
\galaxy\ compares well with the high-luminosity galaxies from the
SAURON sample \citep{Cappellari2007} in that its ellipticity requires
an anisotropic velocity distribution and is not due to rotation.

\begin{figure}
  \centering
  \resizebox{0.9\hsize}{!}{\includegraphics{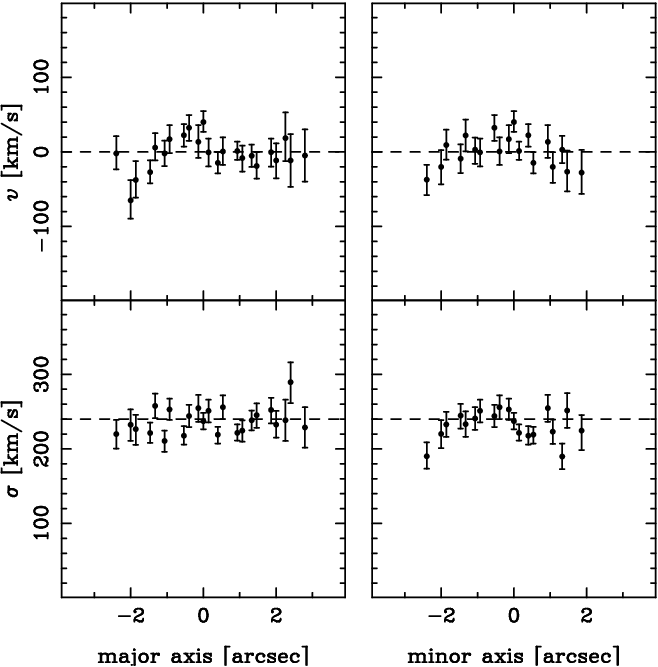}}
  \caption{Velocity (top panels) and velocity dispersion (bottom
    panels) as a function of position along the major (left-hand
    panels) and minor (right-hand panels) axes. Spectra from fibres
    within $1\,$arcsec of the respective axis were used here.}
  \label{fig:kinematic_profiles}
\end{figure} 

The velocity dispersion is nearly constant across the field as one
would expect for a nearly isothermal total mass distribution.
Fig.~\ref{fig:kinematic_profiles} shows the velocity and velocity
dispersion as a function of position along the major and minor axes
(fibres within $1\,\mathrm{arcsec}$ of the respective axis were
used). The major axis velocity cut shows a hint of the S-shape
expected from rotation around the minor axis, although at a very low
level. The velocity dispersion along the major axis is perfectly flat,
whereas the minor axis cut shows a slight decline outwards.

To quantify this, we perform a linear fit to the velocity dispersion
profile as a function of the elliptical radius $\mu^{2} = x^{2}+
y^{2}/\qiso^{2}$, with $\qiso = 0.77$ obtained by ellipse fitting to
the light distribution. The linear slope is
$\mathrm{d}\sigma/\mathrm{d}\mu = (-7.3\pm4.5)\, \mathrm{km\,s^{-1}\,
  arcsec^{-1}}$, consistent with being fully isothermal.

\section{Joint self-consistent gravitational lensing and stellar
  dynamics analysis }
\label{sec:analysis}

In this section, we combine the information from the gravitationally
lensed image with the surface brightness distribution from
\textit{HST} observations and the projected velocity moment maps
derived from VLT observations of the lens galaxy {\galaxy} in order to
constrain the total mass density profile of this galaxy. The analysis
is carried out by making use of the {\dynlen} algorithm, presented in
BK07: we refer to that paper for a detailed description of the method.

\begin{figure*}
  \centering
  \resizebox{0.72\hsize}{!}{\includegraphics[angle=-90]{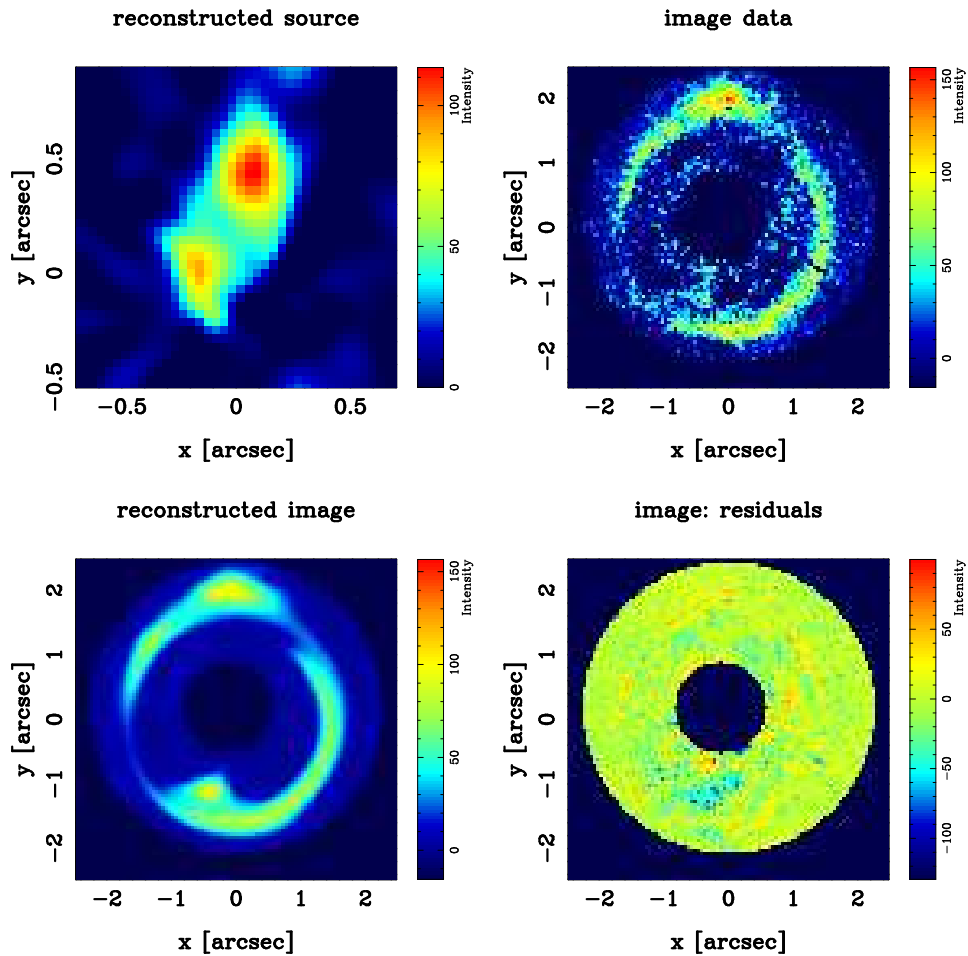}}
  \caption{Best model lens image reconstruction for the system
    {\galaxy}. From the top left-hand to bottom right-hand panel:
    reconstructed source model; \textit{HST}/ACS data showing the
    lensed image after subtraction of the lens galaxy; lensed image
    reconstruction; residuals.}
  \label{fig:LENcomp}
\end{figure*}

\subsection{Mass model and overview of the joint analysis}
\label{ssec:ana:model}

The key idea of a self-consistent joint analysis is to adopt the same
total gravitational potential $\Phi$ for both the gravitational
lensing and the stellar dynamics modelling of the data. As shown in
BK07, these two modelling problems, although different from a physical
point of view, can be expressed in an analogous way as a single set of
coupled (regularized) linear equations. For any given choice $\veceta$
of the non-linear potential parameters, the equations can be solved
non-iteratively to simultaneously obtain as the best solution for the
chosen potential model the unlensed source surface brightness
distributions and the weights of the elementary stellar dynamics
building blocks \citep[e.g.\ orbits or two-integral components,
TICs,][]{Schwarzschild1979, Verolme-deZeeuw2002}.  This linear
optimization scheme is consistently embedded in the framework of
Bayesian statistics. As a consequence, it is possible to objectively
assess the probability of each model by means of the evidence merit
function and, therefore, to rank different models
\citep[see][]{MacKay1992, MacKay1999, MacKay2003}. In this way, by
maximizing the evidence, the set of non-linear parameters $\veceta$
which corresponds to the best potential model, given the data, can be
recovered.

\begin{figure*}
  \centering
  \resizebox{0.78\hsize}{!}{\includegraphics[angle=-90]{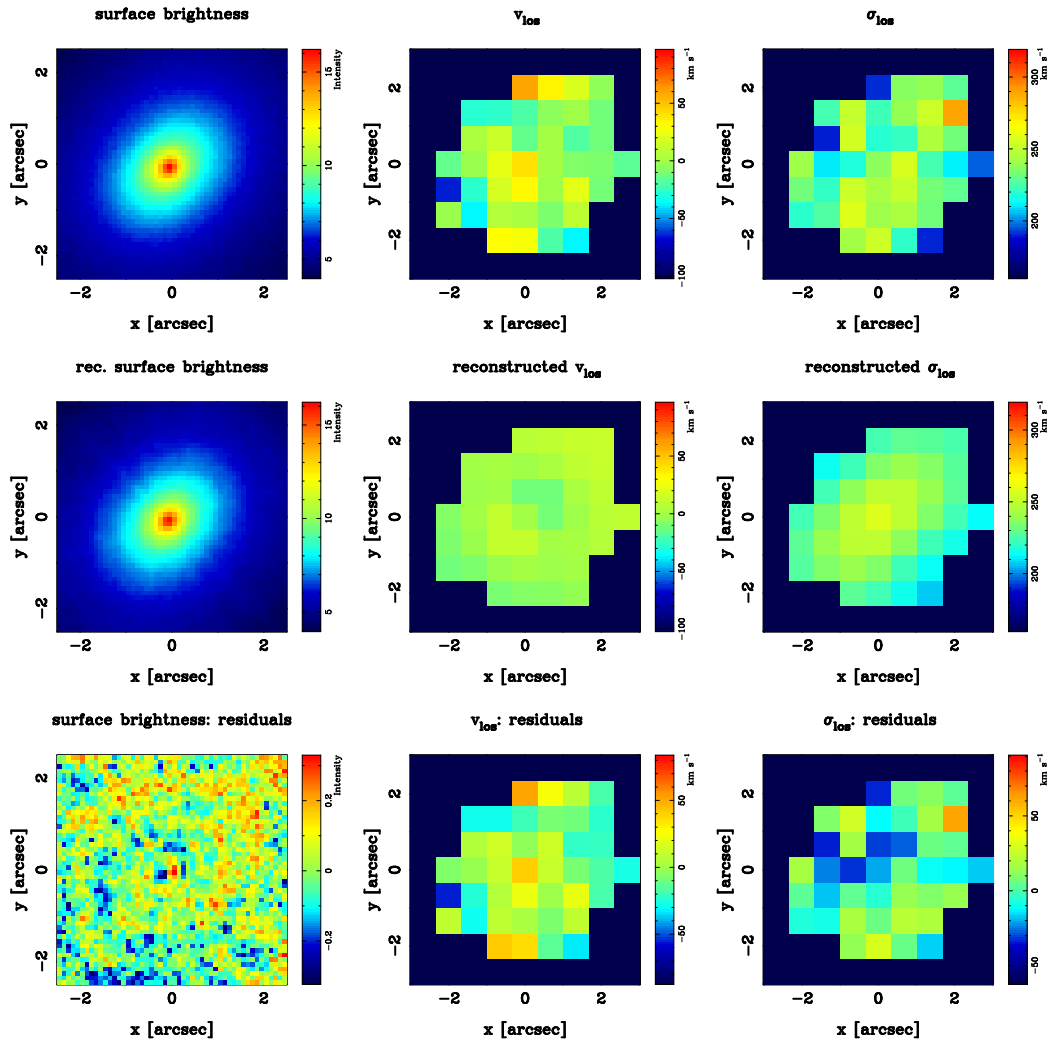}}
  \caption{Best dynamical model for the galaxy {\galaxy}. First row:
    observed surface brightness distribution, projected line-of-sight
    velocity and line-of-sight velocity dispersion. Second row:
    corresponding reconstructed quantities for the best model. Third
    row: residuals. The low-level ripples in the surface brightness
    residuals are due to the discrete nature of the TICs.}
  \label{fig:DYNcomp}
\end{figure*}

While the method is completely general, its practical implementation
(in the {\dynlen} algorithm) is more restricted to make it
computationally efficient and applies specifically to axisymmetric
potentials, $\Phi(R,z)$, and two-integral DFs $f = f(E, \Lz)$. Under
these assumptions, the dynamical model can be constructed by making
use of the fast BK07 numerical implementation of the two-integral
Schwarzschild method developed by \citet{Cretton1999} and
\citet{Verolme-deZeeuw2002}, whose building blocks are not orbits (as
in the classical Schwarzschild method) but TICs.\footnote{A TIC can be
  visualized as an elementary toroidal system, completely specified by
  a particular choice of energy $E$ and axial component of the angular
  momentum $\Lz$. TICs have simple $1/R$ radial density distributions
  and analytic unprojected velocity moments, and by superposing them
  one can build $f(E, \Lz)$ models for arbitrary spheroidal potentials
  \citep[cf.][]{Cretton1999}: all these characteristics contribute to
  make TICs particularly valuable and `inexpensive' building blocks
  when compared to orbits.}  Among the outcomes of the joint analysis
presented in this section is the weights map of the TIC superposition,
which can be related to the weighted two-integral DF (see BK07).

As a first-order model and in order to allow a straightforward
comparison with the results of the preliminary analysis of
\citet{Koopmans2006}, we adopt as the total mass density distribution
of the galaxy a power law stratified on axisymmetric homoeoids:
\begin{equation}
  \label{eq:rho}
  \rho(m) = \frac{\rho_{0}}{m^{\gamma'}}, \quad 0 < \gamma' < 3,
\end{equation}
where $\rho_{0}$ is a density scale, $\gamma'$ will be referred to as
the (logarithmic) slope of the density profile, and
\begin{equation}
  \label{eq:m}
  m^2 = \frac{R^2}{a^2} + \frac{z^2}{c^2} 
  = \frac{R^2}{a^2} + \frac{z^2}{a^2 q^2} ,
\end{equation}
with $a \ge c$ and $c/a \equiv q$.

The (inner) gravitational potential associated with a homoeoidal
density distribution $\rho(m)$ is given by Chandrasekhar's formula
(\citealt{Chandrasekhar1969}; see also e.g.\
\citealt{Merritt-Fridman1996} and \citealt{Ciotti-Bertin2005}). We
find, for $\gamma' \ne 2$,
\begin{equation}
  \label{eq:pot}
  \Phi(R,z) = - \Phi_{0} \int_{0}^{+\infty} 
  \frac{[1/(\gamma'-2)]\,{\mt}^{2-\gamma'}}
  {(1+\tau) \sqrt{q^2 + \tau}}  \,\mathrm{d} \tau\;,
\end{equation}
where $\Phi_{0} = 2 \pi G q a^2 \rho_{0}$ and
\begin{equation}
  \label{eq:mt}
  \mt^{2} = \frac{R^2}{a^2 (1+\tau)} + \frac{z^2}{a^2 (q^2+\tau)} .
\end{equation}
For $\gamma' = 2$ we have
\begin{equation}
  \label{eq:pot.2}
  \Phi(R,z) = - \Phi_{0} \int_{0}^{+\infty} 
  \frac{\log (1/\mt)}
  {(1+\tau) \sqrt{q^2 + \tau}}\,  \mathrm{d} \tau .
\end{equation}

There are three free non-linear parameters in the potential to be
determined via the evidence maximization: $\Phi_{0}$ (or equivalently,
through equation [B4] of BK07, the lens strength $\talp$), the slope
$\gamma'$ and the axial ratio $q$. One can also easily include a core
radius $R_{\mathrm{s}}$ by modifying the definition of the homoeoidal
radius $m$, but for the purpose of the present analysis it was kept
fixed to a negligibly small value. In addition to the previously
mentioned parameters, there are four additional parameters which
determine the geometry of the observed system: the position angle
$\PA$, the inclination $i$ and the coordinates of the centre of the
lens galaxy with respect to the sky grid.

The {\dynlen} algorithm is run on the data sets (and the corresponding
covariance matrices) described in the previous sections.
\begin{enumerate}
\item The lensed image surface brightness distribution is defined on a
  $100 \times 100$ grid ($1~\mathrm{pixel} = 0.05\,\mathrm{arcsec}$)
  on which the innermost and outermost low-$S/N$ regions have been
  masked out, and the source is reconstructed on a $40 \times 40$
  grid. To include the effects of seeing in the modelling, the lensing
  matrix is blurred with a blurring operator which takes into account
  the \textit{HST}/ACS \textit{F814W} point spread function (PSF)
  obtained with \textsc{tiny tim} \citep{Krist1993}.
\item The lens galaxy surface brightness distribution is given on a
  $50 \times 50$ grid ($1~\mathrm{pixel} = 0.10\,\mathrm{arcsec}$),
  while the kinematics data are given on a $9 \times 9$ grid
  ($1~\mathrm{pixel} = 0.67\,\mathrm{arcsec}$), but masking out all
  the points with $S/N$ less than $8$. The dynamical modelling employs
  $100$ TICs, each one populated with $\nTIC = 1 \times 10^{5}$
  particles. The TICs are selected by considering $\nE = 10$ elements
  logarithmically sampled in the radius $\Rc$ of the circular orbit
  with energy $E_{\mathrm{c}} = E(\Rc)$, and $\nLz = 5$ elements
  linearly sampled in angular momentum, mirrored for negative
  $\Lz$. The PSF is modelled as Gaussians with $0.10$~and
  $0.90\,\mathrm{arcsec}$ FWHM for the two grids, respectively.
\end{enumerate}

A curvature regularization (as described in \citealt{Suyu2006} and
appendix~A of BK07) is adopted for both the gravitational lensing and
the stellar dynamics reconstructions. As discussed in BK07, the initial
guess values of the hyperparameters (which set the level of the
regularization) are chosen to be quite large, since the convergence to
the maximum is faster when starting from an overregularized system.
Moreover, in order to further speed up the reconstruction, the
position angle $\PA$ and the coordinates of the lens galaxy centre
$\vec{\xi}_{\mathrm{c}} = (x_{\mathrm{c}}, y_{\mathrm{c}})$ are fairly
accurately determined by means of a preliminary optimization run and
subsequently kept fixed or allowed to vary only within a limited range
around the determined values. The position angle $\PA = 135\fdg5$
obtained in this way is very close to both the observed value for the
light distribution $\vartheta_{\star} = 126\fdg5$ and the position
angle $\vartheta_{\mathrm{SIE}} = 136\fdg2$ obtained for the singular
isothermal ellipsoid lensing model in \citet{Koopmans2006}.

\subsection{Results}
\label{ssec:ana:results}

\begin{table}
  \centering
  \begin{minipage}{0.85\hsize}
    \caption{{\galaxy}: non-linear parameters for the best power-law
      model $\rho \propto m^{-\gamma'}$ (equation~\ref{eq:rho}).}
    \smallskip
  \begin{tabular*}{\hsize}{@{}@{\extracolsep{\fill}}c c c }
    \hline
      \noalign{\smallskip}
      Model: power law & & $\model_{\mathrm{best}}$ \\
      \noalign{\smallskip}
      \hline
      \noalign{\smallskip}
                 & $i$             & $67\fdg8$ \\
      Non-linear & $\talp$         & $0.468$   \\
      parameters & $\gamma'$       & $2.061$  \\
                 & $q$             & $0.739$   \\
      \noalign{\smallskip}
      \noalign{\smallskip}
                       & $\log \lamlen$  & $-2.344$  \\
      \noalign{\smallskip}
      Hyperparameters  & $\log \lamx$    & $-0.023$  \\
      \noalign{\smallskip}
                       & $\log \lamy$    &  $1.810$  \\
      \noalign{\smallskip}
      \noalign{\smallskip}
               & $\evid_{\mathrm{len}}$ & $-78\,480$ \\
      Evidence & $\evid_{\mathrm{dyn}}$  & $-14\,558$ \\
               & $\evid_{\mathrm{tot}}$ & $-93\,038$ \\
      \noalign{\smallskip}
      \hline
    \end{tabular*}
    
  \end{minipage}
  \label{tab:results}
\end{table}

The non-linear parameters of the best-fitting model obtained from the
combined gravitational lensing and stellar dynamics reconstruction are
presented in Table~\ref{tab:results}, together with the best-recovered
hyperparameters and the values of the evidence.  The recovered
inclination angle is $i = 67\fdg8$. The lens strength, proportional to
the normalization constant of the potential $\Phi_{0}$ (see appendix~B
of BK07), is $\alpha_{0} = 0.468$. The logarithmic slope and the axial
ratio of the total density distribution of the power-law model are
$\gamma' = 2.061$ and $q = 0.739$, respectively.

\begin{figure}
  \centering
  \resizebox{0.82\hsize}{!}{\includegraphics[angle=-90]{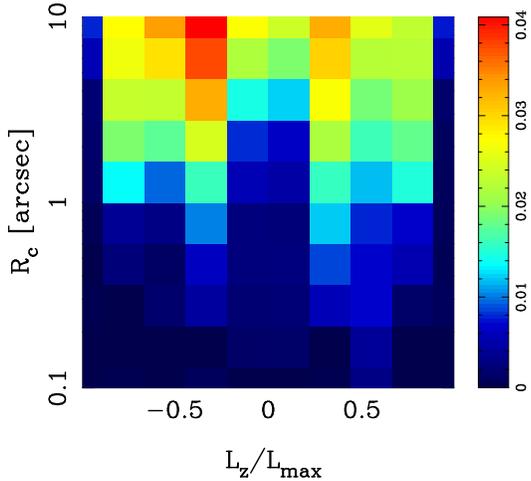}}
  \caption{Reconstruction of the weighted two-integral DF of the lens
    galaxy {\galaxy} obtained from the best-fitting model. Each pixel
    in the two-integral space gives the relative contribution of the
    corresponding TIC to the total mass of the modelled system. Here,
    $L_{\mathrm{max}}$ indicates the maximum value of the angular
    momentum along the $z$ axis: $L_{\mathrm{max}} = \Rc \vc$, where
    $\Rc$ is the circular radius and $\vc(\Rc)$ is the circular
    velocity corresponding to a total energy $E(\Rc) =
    \Phi_{\mathrm{eff}}(\Rc, 0) = \Phi(\Rc,0) + \vc^{2}/2$.}
  \label{fig:DFcomp}
\end{figure}

The reconstructed observables corresponding to the best-fitting model
are shown and compared to the data in Figs~\ref{fig:LENcomp}
and~\ref{fig:DYNcomp} for, respectively, gravitational lensing and
stellar dynamics. Fig.~\ref{fig:LENcomp} also displays the
reconstructed source model, while the reconstruction of the weighted
two-integral DF of the lens galaxy is given in Fig.~\ref{fig:DFcomp}.

The presence of an external shear was also examined, giving results
consistent with zero. Furthermore, in order to assess the possible
effect of a faint galaxy located at $(2\farcs55, 2\farcs10)$
north-west of the centre of the lens galaxy on the sky
(Fig.~\ref{fig:HST_images}), a singular isothermal sphere (SIS)
centred on the corresponding position was added to the model, but the
best reconstruction unambiguously excluded any relevance of this
component (its lens strength is not significant).

As a sanity check, we also performed the reconstruction with a
different set-up for the dynamics, that is, employing a larger number
of TICs ($\nE = 18$ and $\nLz = 9$) or increasing the number of
populating particles to $\nTIC = 12 \times 10^{5}$ while maintaining
the same number of TICs. In both cases, the obtained values for the
non-linear parameters do not differ significantly from the ones
presented in Table~\ref{tab:results}. There is an improvement in the
total evidence mainly because of the slightly better fit to the
surface brightness distribution.

The most remarkable result of the reconstruction is the logarithmic
slope $\gamma' \simeq 2$, making the total density profile very close
to isothermal. This is in good agreement with the previous finding of
$\gamma' \simeq 1.9$ obtained by \citet{Koopmans2006} for the same
system.  The fact that \citet{Koopmans2006} achieved robust results
with a simpler and not fully self-consistent joint lensing and
dynamics analysis (in which they only used the average
$\sigma_{\mathrm{los}}$ from the SDSS fibre spectroscopy as a
constraint for the dynamics and solved the spherical Jeans equations)
is a strong indication of the effectiveness of the combined analysis
in pinning down the characteristics of the total potential.

It should be stressed that the role of the kinematic constraints is
crucial in breaking the ample degeneracies that would arise from a
reconstruction limited to the lensing data: in particular, the fairly
constant line-of-sight velocity dispersion map unyieldingly rules out
values for the slope which are significantly far from isothermal, but
which would turn out to be perfectly reasonable (even based on the
quantitative evidence) if lensing alone was considered.

The lens galaxy surface brightness distribution together with the
line-of-sight projected velocity map (which shows a slight indication
of rotation of the system) provide important information on the
flattening and the inclination of the system, although these
quantities remain the most difficult to disentangle. The weak (of the
order of a few per cent) but discernible pattern in the residuals of
the surface brightness reconstruction can be ascribed to the effect of
the discreteness of the adopted TIC library, but could also be an
indication of a slight offset in the orientation between the luminous
matter distribution and the total potential: when the reconstruction
is limited to lensing alone, the obtained position angle is typically
a few degrees larger than $\PA$ found for the joint reconstruction.

Much information about the best reconstructed dynamical model is
contained in the weighted two-integral DF presented in
Fig.~\ref{fig:DFcomp} (refer to appendix~D of BK07 for a precise
definition of the weights $\gamma_j$ which are the quantity shown in
this plot). Since all the original TICs have equal mass by
construction, each pixel in the integral space provides the relative
contribution in mass of the corresponding TIC. The weighted DF is only
slightly asymmetric around $\Lz/L_{\mathrm{max}} = 0$, which is
reflected in the very low values of the rotation velocity of the best
model system (cf.\ central panel of Fig.~\ref{fig:DYNcomp}). The very
weak indication of counter-rotation which can be detected in the same
panel is also more clearly exhibited in the weighted DF: for the
higher $\Rc$ the TICs with negative $\Lz$ (the red pixels) show larger
values than the corresponding (mirrored) TICs with positive $\Lz$,
while the opposite behaviour is displayed for $\Rc \la 1$.  Given the
low $S/N$, however, we do not place much significance on this possible
counter-rotation, although SAURON has discovered several systems with
such a behaviour \citep{Emsellem2004}.

\subsection{Inferences from the phase-space distribution}
\label{ssec:phase-space_distribution}

From the best reconstructed DF we calculate the axial ratio $\qlight$
of the three-dimensional light distribution, which does not need to
coincide with the axial ratio of the total density distribution
$\rho(m)$ since in our approach mass is not required to follow
light. We adopt the definition
\begin{equation}
  \label{eq:qlight}
  \qlight^{2} = 2 \frac{\int \rho_{\star} z^{2}\,\mathrm{d}V}
                       {\int \rho_{\star} R^{2}\,\mathrm{d}V} ,
\end{equation}
where $R$ and $z$ are the usual cylindrical coordinates and $V$ is the
volume.  This definition has been chosen such that a density
distribution which is a function of the elliptical radius $\ell^{2} =
R^{2} + z^{2}/{q_{\mathrm{ell}}}^{2}$ (e.g.\ a prolate or oblate
ellipsoid of axial ratio $q_{\mathrm{ell}}$) gives $\qlight =
q_{\mathrm{ell}}$. For our best model we obtain an axial ratio
$\qlight = 0.847$, which is rounder than the $q$ of the total density
distribution.  From the observed isophotal two-dimensional axial ratio
$\qiso = 0.77$ (measured by fitting ellipses to the surface brightness
contours of the galaxy, Table~\ref{tab:basic_data}) and from the best
model inclination of $67\fdg8$, one obtains a three-dimensional axial
ratio of the stellar density of $q_{\mathrm{iso,3D}}=0.72$. This does
not coincide with the value of $\qlight$ calculated from the second
moments of the luminosity density of the galaxy
(equation~\ref{eq:qlight}). The disagreement might indicate that the
recovered three-dimensional light distribution is not constant on
ellipsoidal surfaces or that there is a change in ellipticity as a
function of radius.

\begin{figure} 
  \resizebox{0.84\hsize}{!}{\includegraphics[angle=-90]{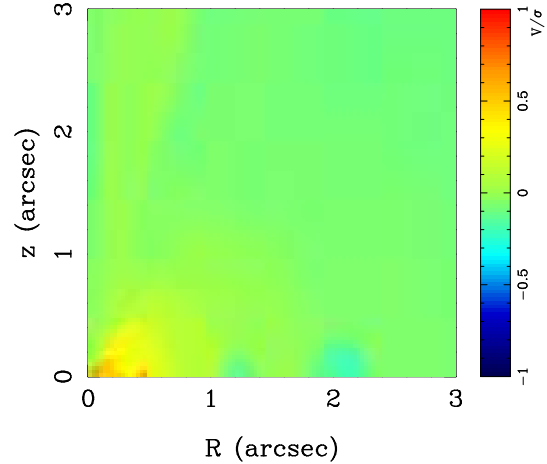}}
  \vspace{0.0cm}
  \caption{Ratio $\vphi/\bar{\sigma}$ of the first and second velocity
    moments for the best reconstructed model of galaxy {\galaxy},
    plotted in the positive quadrant of the meridional plane.}
  \label{fig:VoS}
\end{figure}

Following \citet{Cappellari2007}, we plot in Fig.~\ref{fig:VoS} the
ratio $\vphi/\bar{\sigma}$ in the meridional plane between the mean
rotation velocity around the $z$-axis and the local mean velocity
dispersion $\bar{\sigma}^{2} = (\sigma_{R}^{2} + \sigma_{\varphi}^{2}
+ \sigma_{z}^{2})/3$. This quantity provides an indication of the
importance of the rotation with respect to the random motions for each
position $(R,z)$ in the galaxy. From the analysis of
Fig.~\ref{fig:VoS}, {\galaxy} appears to be an overall slow rotator,
although rotation seems to play a more significant role in the inner
regions. Two small regions characterized by slight counter-rotation
are located around the equatorial plane ($z\!=\!0$). Due to the
assumption of a two-integral DF, the cross-section of the velocity
dispersion ellipsoids would be circular for every position on the
meridional plane.

\begin{figure}
  \centering
  \resizebox{0.78\hsize}{!}{\includegraphics[angle=-90]{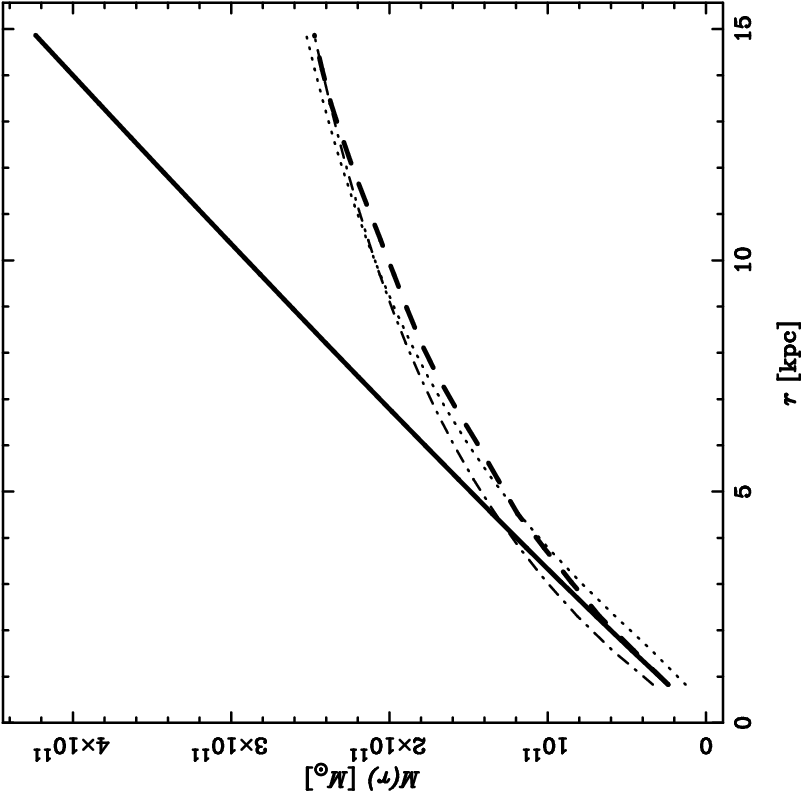}}\\
  \vspace{2ex}
  \resizebox{0.78\hsize}{!}{\includegraphics[angle=-90]{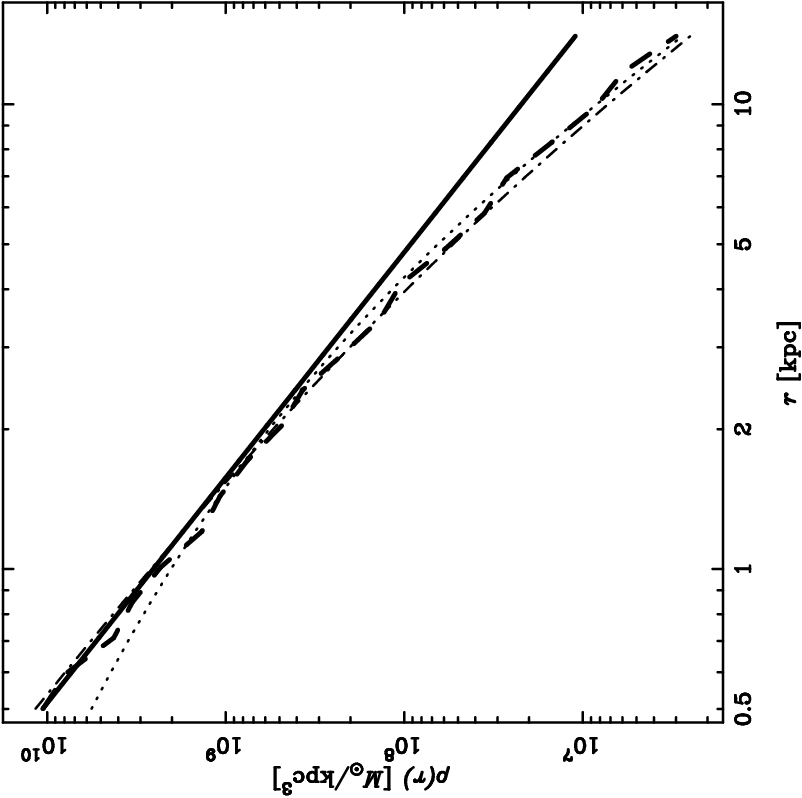}}
  \caption{Upper panel: mass enclosed within a radius $r$. The thick
    solid line shows the total mass of the galaxy as a function of $r$
    obtained from the best reconstructed model. The thick dashed line
    shows the same quantity for the luminous component obtained from
    the reconstructed DF and rescaled to the value which maximizes the
    contribution of luminous over dark matter (`maximum bulge'). As a
    comparison, the dash--dotted and dotted lines show the quantity
    $M(r)$ for, respectively, a Jaffe and a Hernquist profile,
    normalized to the total luminous mass. Lower panel: logarithm of
    circularized mass density as a function of radius. The line styles
    have the same meaning as above.}
  \label{fig:profile}
\end{figure}

Another helpful way to quantify the anisotropy of an axisymmetric galaxy is
provided by the three global anisotropy parameters defined in
\citet{BT87} and \citet{Cappellari2007}:
\begin{equation}
  \label{eq:AP:beta}
  \beta \equiv 1 - \frac{\Pi_{zz}}{\Pi_{RR}}, 
\end{equation}
\begin{equation}
  \label{eq:AP:gamma}
  \gamma \equiv 1 - \frac{\Pi_{\varphi\varphi}}{\Pi_{RR}}, 
\end{equation}
\begin{equation}
  \label{eq:AP:delta}
  \delta \equiv 1 - \frac{2 \Pi_{zz}}{\Pi_{RR} + \Pi_{\varphi\varphi}} = 
  \frac{2 \beta - \gamma}{2 - \gamma},
\end{equation}
where 
\begin{equation}
  \label{eq:AP:PI}
  \Pi_{kk} = \int \rho \sigma^{2}_{k}\, \mathrm{d}^{3}x 
\end{equation}
and $\sigma_{k}$ denotes the velocity dispersion along the direction
$k$ at any given location in the galaxy. For nearly spherical systems
it can be convenient to consider the anisotropy parameter $\beta_{r}$
defined as
\begin{equation}
  \label{eq:AP:beta_r}
  \beta_{r} \equiv 1 - \frac{\Pi_{tt}}{\Pi_{rr}}
\end{equation}
in the spherical coordinates $(r, \theta, \varphi)$. Here $\Pi_{tt} =
(\Pi_{\theta \theta} + \Pi_{\varphi \varphi})/2$.

For a two-integral DF, as mentioned above, $\sigma_{R}^{2} =
\sigma_{z}^{2}$ everywhere, which implies $\Pi_{RR} =
\Pi_{zz}$. Therefore, the value of $\beta$ is always zero, while
$\gamma$ and $\delta$ have opposite signs. Moreover, under this
assumption it is easy to show that $\beta_{r} = \gamma/2$.

Calculating the non-zero anisotropy parameters for the best
reconstructed model, we find $\gamma = -0.32$, which indicates a
slight tangential anisotropy, and $\delta = 0.14$.  The latter value
agrees well with the corresponding value determined directly from the
observed data in Section~\ref{sec:inferences}. The anisotropy
parameters found for {\galaxy} have values which are consistent with
the findings of \citet{Cappellari2007} for their most luminous
galaxies. It should be pointed out, however, that the comparison is
not immediately straightforward, since their use of a three-integral
Schwarzschild orbit-superposition method gives more flexibility to the
models.

We compared the radial profile of the total mass density of the galaxy
(given by equation~\ref{eq:rho} by inserting the best-fitting model
parameters) with the density of the stellar component alone, as
obtained directly from the reconstructed DF. The axisymmetric density
distribution was circularized by averaging over spherical shells. The
stellar density profile has been normalized by using as a constraint
the value $\Meff \simeq 2.0 \times 10^{11} M_{\sun}$ for the luminous
mass inside the effective radius, which is obtained from the
observations when assuming an average $M/L$ of
$5.2~(M/L)_{\sun,B}$. This `maximum bulge' rescaling (analogous to the
maximum disc approach for spiral galaxies) maximizes the contribution
of the luminous component with respect to the total mass. The density
and the enclosed mass as a function of radius are plotted in
Fig.~\ref{fig:profile}, together with the same functions for the
spherical Jaffe and Hernquist profiles \citep{Jaffe1983,
  Hernquist1990} normalized to the total mass $2 \Meff$. While very
simple, these models appear to provide a reasonably good approximation
to the light distribution. From the diagrams, one can note that the
total mass density distribution follows the light in the inner regions
up to a distance of about 5\,kpc, where the contribution of the dark
matter component starts to become significant: at 5\,kpc the
non-visible mass represents about 15~per cent of the total mass,
rising to $\sim 30$~per cent at 10\,kpc (the distance for which the
enclosed stellar mass approximately equals $\Meff$, i.e.~the distance
corresponding to the unprojected effective radius) and $\sim 40$~per
cent at 15\,kpc.

The spatial coverage of the data might appear quite limited when it
comes to the study of the total density profile at radii of the order
of 10\,kpc, since the Einstein radius $\REin \simeq
2.6\,\mathrm{kpc}$, the effective radius $\Reff$ is of the order of
8\,kpc (which however becomes $\sim 10$ kpc when unprojected) and the
integral-field kinematics does not extend much farther than
3\,kpc. However, one should note that a fair amount of information
comes also from those distant regions of the galaxy that are seen in
projection along the line of sight.  In order to verify this, we
consider the intersection between a sphere of radius $r$, centred on
the galaxy, and a cylinder oriented along the line of sight which has
a radius equal to, respectively, the Einstein radius and the effective
radius.  In the first case, it is found that for $r = \REin$ the mass
enclosed within this volume corresponds to approximately 75~per cent
of $\MEin$ (the total mass within the Einstein radius). In other
words, one-fourth of the contribution to the gravitational lensing
comes from matter which is located farther than 2.6\,kpc but falls, in
projection, within the Einstein radius.  In the second case, we obtain
that for $r = \Reff$ the enclosed light is about 77~per cent of
$\Leff$, that is, again almost one-fourth of the luminosity enclosed
within the cylinder of radius $\Reff$ comes from regions at a radial
distance $r > \Reff$ from the centre of the galaxy.  Any changes in
the density distribution at radial distances larger than $\REin$ or
$\Reff$ which still influence the data model inside those projected
radii will have an effect which is reflected statistically in the
evidence (or in the likelihood), allowing to discriminate between
different models. Clearly, the effect becomes progressively weaker
with increasing radial distances.

We also explored a constant-$M/L$ model, by considering a double
power-law density distribution which approximates the reconstructed
luminous density profile of Fig.\,\ref{fig:profile}: this density
profile is nearly isothermal in the inner regions but the slope
becomes significantly steeper ($\gamma'_{\mathrm{out}} \sim 3.3$)
outwards. The $M/L$ is $5.9 \, (M/L)_{\sun,B}$. We find that the
constant-$M/L$ model, while able to reproduce the data, fits the
dynamics somewhat worse: the $\chi^{2}$ per pixel for the velocity
dispersion is $\sim 2.2$, higher than in the case of the single
power-law model ($\chi^{2}$ per pixel $\sim 1.4$). For the lensing and
the surface brightness distribution, instead, the values of $\chi^{2}$
per pixel of the two models are comparable (with the double power law
being just a few per cent higher). In conclusion, under the considered
assumptions, the nearly isothermal single power-law model is slightly
preferred by the data over a constant-$M/L$ model with no dark matter.
It should be noted, however, that the difference between the
constant-$M/L$ model and the power-law model (for which mass does not
need to follow light) is relatively small. It is possible that much or
all of this discrepancy could disappear if one allowed for more
sophisticated three-integral dynamical models.

\subsection{Uncertainties}
\label{ssec:ana:errors}

\begin{table*}
  \centering
  \begin{minipage}{0.82\hsize}
  \caption{Summary of the model uncertainties, determined by
    considering {\Nrealiz} random realizations of the best model data
    sets. The reconstructed parameters obtained for the best model
    $\model_{\mathrm{best}}$ are also reported for
    comparison.  \label{tab:stat} }
  \smallskip
  \begin{tabular*}{\hsize}{@{}@{\extracolsep{\fill}}c c c c c c c c }
    \hline
    \noalign{\smallskip}
    & & Median & Mean & 68 per cent confidence interval & 95 per cent
    confidence interval & $\model_{\mathrm{best}}$ \\
    \noalign{\smallskip}
    \hline
    \noalign{\smallskip}
    & $i$       & $66\fdg1$ & $65\fdg2$  & [$60\fdg0$,  $68\fdg9$] & [$55\fdg1$,  $75\fdg8$]  & $67\fdg8$ \\
    Non-linear & $\talp$   & 0.472          & 0.472           & [0.467, 0.475]                    & [0.463, 0.479]                     & 0.468  \\
    parameters & $\gamma'$ & 2.061          & 2.046           & [1.996, 2.085]                    & [1.894, 2.142]                     & 2.061  \\
    & $q$       & 0.739          & 0.730           & [0.688, 0.760]                    & [0.657, 0.802]                     & 0.739  \\
    \noalign{\smallskip}
    \hline
  \end{tabular*}
\end{minipage}
\end{table*}

\begin{figure*}
  \centering
  \resizebox{0.88\hsize}{!}{\includegraphics[angle=-90]{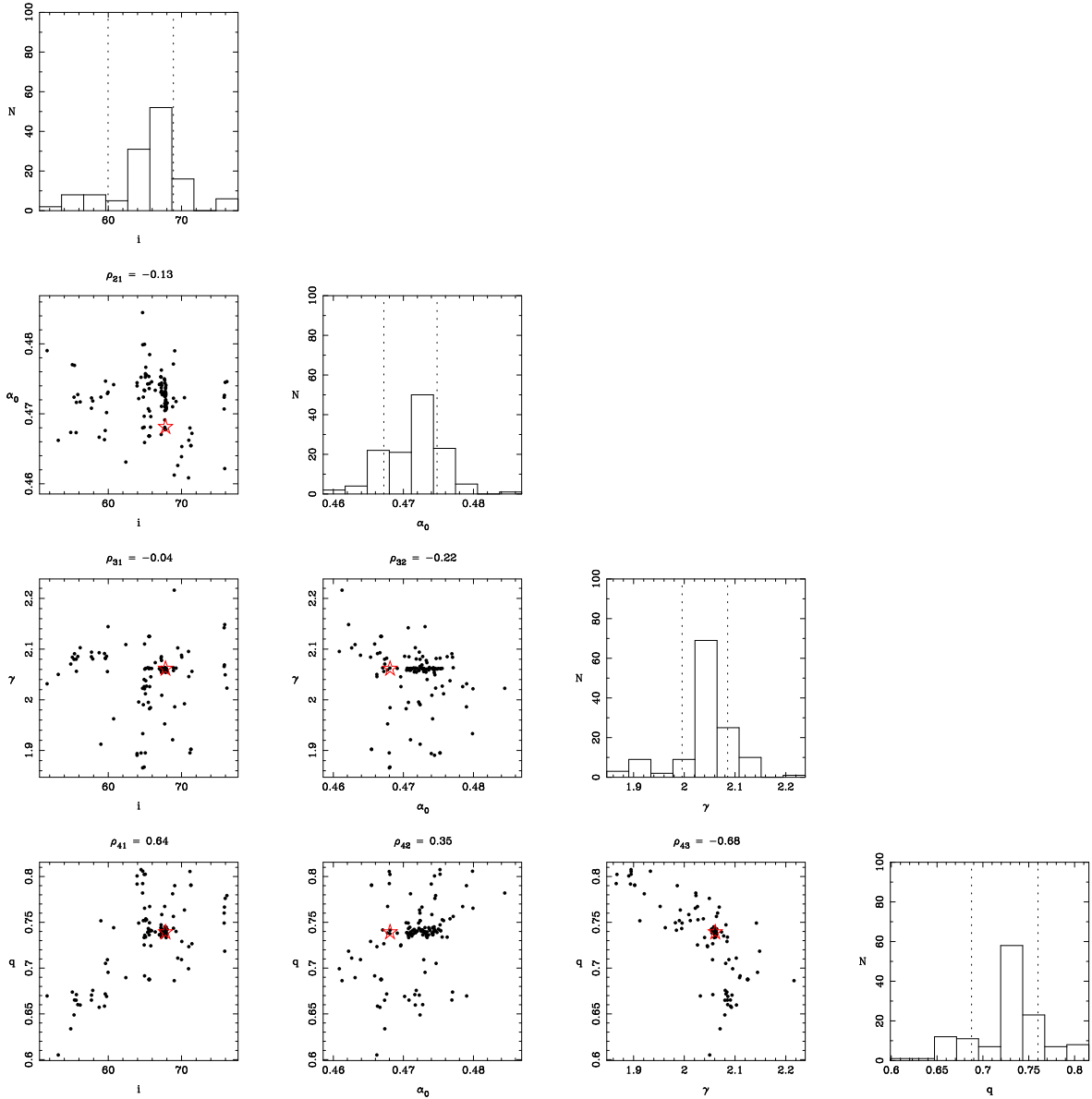}}
  \caption{Graphical visualization of the lower triangle of the
    (symmetric) correlation matrix for the parameters recovered from
    the non-linear reconstruction of {\Nrealiz}~random realizations of
    the best model data sets. The panels on the main diagonal present
    the distribution of the four non-linear parameters $i$,
    $\alpha_{0}$, $\gamma'$ and $q$ over the realizations. In the
    remaining panels the non-linear parameters are plotted two by two
    against each other; for each panel, the corresponding value
    $\rho_{ij}$ of the correlation matrix is also indicated. The
    dotted vertical lines indicate the 68~per cent confidence
    interval. The red stars indicate the locations of the parameters
    obtained from the best model reconstruction
    $\model_{\mathrm{best}}$ (cf.\ Table~\ref{tab:results}).}
  \label{fig:corr}
\end{figure*}

The model uncertainties, that is, the scatter in the recovered
non-linear parameters $i$, $\alpha_{0}$, $\gamma'$ and $q$, were
determined by considering {\Nrealiz} random realizations of the
datasets and rerunning the {\dynlen} algorithm for each of
them. Table~\ref{tab:stat} summarizes the results of the statistical
analysis and shows the 68~and 95~per cent confidence intervals for the
four non-linear parameters. Fig.\,\ref{fig:corr} shows a graphical
visualization of the correlation matrix, displaying the parameters
plotted against each other, while the histograms on the main diagonal
present the distribution of the four parameters over the
realizations. These distributions, which provide the error bars for
the best-fitting model parameters, can immediately be seen to be
generally non-Gaussian.

As highlighted in BK07, there appears to be a partial degeneracy
between the inclination~$i$ and the axial ratio~$q$, which is however
limited only within a certain interval in the parameters. The lens
strength $\alpha_{0}$ is quite tightly constrained, and the
logarithmic slope $\gamma'$ spans over an interval of values all very
close to the isothermal $\gamma' = 2$ case.


\section{Breaking degeneracies: oblateness and inclination}
\label{sec:degeneracies}

When combining stellar kinematic and gravitational lensing data of an
early-type galaxy, even under the simplifying assumption of spherical
symmetry, it has been shown that the degeneracies between mass,
orbital anisotropy and total density slope can effectively be broken
\citep[e.g.][]{Treu-Koopmans2002a, Koopmans2003, Koopmans-Treu2003,
  Treu-Koopmans2004, Koopmans2006}.
More recently, the theoretical basis of this methodology was made
self-consistent and extended to include flattened axisymmetric mass
distributions and the modelling of the stellar two-integral
phase-space DF $f(E,L_{z})$ (BK07). It was shown that under the
assumptions of axisymmetry and two-integral DFs not only the
degeneracy between mass, mass slope and anisotropy can be broken, but
also that between the flattening (i.e.\ oblateness) of the mass
distribution and its inclination with respect to an observer. The
latter degeneracy poses a restriction on most current dynamical
modelling efforts, which often have to assume some (range of)
inclination \citep[e.g.][]{Emsellem2007, Cappellari2007}.

By combining the information from lensing and dynamics we have found a
well-defined mass slope (consistent with \citealt{Koopmans2006}),
inclination and oblateness for the early-type lens galaxy
SDSS\,J2321$-$097 (see e.g.\ Fig.~\ref{fig:corr}). In this section, we
show why the degeneracy between oblateness and inclination can, in
principle, be broken. As in all of this paper, we assume axisymmetry
and restrict the phase-space density distribution to be a function of
the two classical integrals of motion, $E$ and $\Lz$ (these
assumptions hold well in the system considered in this paper).
In this case, it can be shown from symmetry arguments \citep{BT87}
that at any point in the galaxy the two-dimensional cut through the
stellar velocity ellipsoid in the meridional plane $(R, z)$ is round
(i.e.\ $\sigma_{R}=\sigma_{z}$). For an observer located in the
meridional plane, the line-of-sight component of the stellar velocity
dispersion from any given point on the same plane will therefore
appear independent of inclination.
The luminosity-weighted integral of the stellar velocity dispersion
over the entire meridional plane is then also independent of
inclination.  Because an observer is always situated in the meridional
plane spanned by himself and the minor axis of an axisymmetric galaxy,
we state that:
\begin{quote}
  \itshape The surface brightness weighted integral over the
  line-of-sight stellar velocity dispersion along the minor axis of an
  axisymmetric galaxy with a DF that depends only
  on energy and angular momentum, $f(E, L_{z})$, is independent of
  inclination. This also holds in the case of streaming motions which
  are perpendicular to the minor axis.
\end{quote}

This can be formally proved as follows. The luminosity-weighted
stellar velocity dispersion integrated over the meridional plane of a
galaxy with inclination $i=0$ is given by
\begin{equation}
  \langle \sigma_{\mathrm{mp}}^2 \rangle (i=0) 
  = \frac{\int_{-\infty}^{\infty} \int_{-\infty}^{\infty}
    \sigma^2_{\mathrm{los}}(R,z)\,\rho_{\star}(R,z)\,
    \mathrm{d}R\,\mathrm{d}z} 
  {\int_{-\infty}^{\infty} \int_{-\infty}^{\infty} \rho_{\star}(R,z)\,
    \mathrm{d}R\,\mathrm{d}z} \,.
\end{equation}
Under a rotation, denoted by the orthogonal rotational matrix
$\mat{R}(i)$ with $\det \mat{R}(i) = \det \mat{R}^{-1}(i) = 1$, the
coordinate system can be transformed as $\vec{x}'=\mat{R}(i) \vec{x}$
with $\vec{x}\equiv (R,z)$ and $\vec{x}'\equiv (R',z')$. The above
equation can then, after coordinate transformation, be written as
\begin{equation}
  \langle \sigma_{\mathrm{mp}}^2 \rangle (i) 
  = \frac{\int_{-\infty}^{\infty} \int_{-\infty}^{\infty} 
    \sigma^2_{\mathrm{los}}(R',z')\,
    \rho_{\star}(R',z')\,\det \mat{R}^{-1}(i)\,
    \mathrm{d}R'\,\mathrm{d}z'} 
  {\int_{-\infty}^{\infty} \int_{-\infty}^{\infty}
    \rho_{\star}(R',z')\,\det \mat{R}^{-1}(i)\, \mathrm{d}R'\,\mathrm{d}z'}.
\end{equation}
Because rotation does not alter the line-of-sight velocity dispersion
of a given point in the meridional plane if the potential is
axisymmetric and the DF is only a function of $E$
and $L_z$, it is easy to see that $\sigma^{2}_{\mathrm{los}}(R',z') =
\sigma^{2}_{\mathrm{los}}(R,z)$ for any inclination $i$.  Also the
scalar $\rho_{\star}$ is invariant under rotation, hence
$\rho_{\star}(R',z') = \rho_{\star}(R,z)$.  Given the fact that $\det
\mat{R}^{-1}(i)=1$, it immediately follows that
\begin{equation}
  \langle \sigma_{\mathrm{mp}}^{2} \rangle (i) \equiv \langle
  \sigma_{\mathrm{mp}}^{2}  \rangle (i=0),
\end{equation}
proving the above statement. Hence, $\langle \sigma_{\mathrm{mp}}^2
\rangle (i)$\ effectively becomes a function only of the density
profile, mass and flattening of the mass distribution. Since the
former two are well constrained by gravitational lensing alone and
even better in combination with the stellar kinematics, $\langle
\sigma_{\mathrm{mp}}^2 \rangle (i)$ reduces further to a function of
mostly the oblateness: for a given mass and mass profile, a more
oblate galaxy will have a larger value of $\langle
\sigma_{\mathrm{mp}}^2 \rangle (i)$. Because $\qlight$ and inclination
are restricted by the observed brightness distribution of the galaxy,
and the oblateness by the value of $\langle \sigma_{\mathrm{mp}}^2
\rangle$, one can solve for the inclination and oblateness
simultaneously.

From the IFU data the value of $\langle \sigma_{\mathrm{mp}}^2
\rangle$ can be inferred (because of the limited field of view, we
cannot integrate over the entire minor axis and $\langle
\sigma_{\mathrm{mp}}^2 \rangle$ might therefore still be a weak
function of $i$) and in combination with the observed flattening of
the galaxy brightness distribution, the inclination can be
inferred. It is clear that this can only be done when the stellar
velocity dispersion along the entire minor axis is known and that the
inclination cannot be inferred from a \emph{total} luminosity-weighted
stellar velocity dispersion.

We again emphasize here that the above result only holds in
axisymmetric two-integral situations. To what extent it remains valid
in cases where a third integral of motion is allowed (or axisymmetry
is broken), remains a subject of further study.

\section{Discussion and conclusions}
\label{sec:discussion}

In this paper we have presented the first results from an
integral-field spectroscopic survey of early-type lens galaxies from
SLACS. The combination of integral-field spectroscopy from VIMOS/IFU
mounted on the VLT with high-resolution imaging from \textit{HST}/ACS
has enabled us to conduct the first in-depth study of the structure of
a luminous elliptical galaxy beyond the local Universe, \galaxy\ at
$z=0.0819$.  We have applied a new analysis method that combines the
kinematic and lensing information in a fully self-consistent way and
have shown how this combination breaks some of the degeneracies that
limit the separate application of these two methods.

The galaxy that we have studied here turns out to be a fairly ordinary
elliptical with properties similar to those of local galaxies of
comparable luminosity, such as those studied by the SAURON
collaboration \citep{Emsellem2007, Cappellari2007}.  \galaxy\ is a
slow rotator in the classification of \citet{Emsellem2007} with an
angular momentum parameter of $\lambda_{\mathrm{R}} = 0.075$. The
velocity dispersion map is flat to the limit where we were able to
measure the kinematic parameters reliably. Using the updated estimator
of \citet{Binney2005} for the ratio between systematic and random
velocities, $v/\sigma_{v}$, that fully exploits the information in
integral-field spectroscopic data, we have shown that the ellipticity
of the stellar distribution in \galaxy\ is due to anisotropy of the
velocity distribution rather than rotation, again in line with local
galaxies of comparable luminosity \citep{Cappellari2007}.

We have modelled the galaxy by making use of the {\dynlen} algorithm
which self-consistently combines gravitational lensing and stellar
dynamics under the assumption of axisymmetry and a two-integral DF. We
adopted for the system the total gravitational potential generated by
an axisymmetric power-law mass density distribution of logarithmic
slope $\gamma'$ and axial ratio $q$. The best-fitting model given the
data is obtained in the framework of Bayesian statistics by maximizing
the evidence merit function. The results for the best-fitting model
are summarized as follows.
\begin{enumerate}
\item The logarithmic slope of the total density is $\gamma' =
  2.06^{+0.03}_{-0.06}$ (the error is given within the 68~per cent
  confidence interval), which is very close to (and consistent with)
  an isothermal density distribution.
\item The axial ratio of the total density distribution is $q =
  0.74^{+0.02}_{-0.05}$. Since in our approach mass is not required to
  follow light, this $q$ does not have to coincide with the (average)
  axial ratio of the luminous distribution, which is calculated from
  the reconstructed DF, giving $\qlight = 0.85$ for the best model.
\item The inclination angle of the galaxy is $i =
  67\fdg8^{+1.1}_{-7.8}$. The small error bar shows that inclination
  can be well determined in combination with lensing data.
\item The `maximum bulge' approach, that is, the rescaling of the
  circularized stellar density profile which maximizes the
  contribution of the luminous component to the total density profile,
  prescribes a stellar mass $M_{\mathrm{eff}} \simeq 2.0 \times
  10^{11} M_{\sun}$ inside the effective radius, which corresponds to
  an average $M/L$ of $5.2\,(M/L)_{\sun,B}$. The total mass enclosed
  in the same region is approximately $2.9 \times 10^{11} M_{\sun}$:
  the non-visible matter, therefore, accounts for about 30 per cent of
  the total mass within that three-dimensional radius.
\item The local $\vphi/\bar{\sigma}$ ratio on the meridional plane
  confirms that {\galaxy} as a whole is a slow rotator, with the
  random motions becoming less predominant compared to rotation only
  in the very central regions. The best model yields a global
  anisotropy parameter $\delta = 0.14$, fully consistent with the
  value obtained directly from the data, showing that the galaxy is
  close to an isotropic rotator.  The other global anisotropy
  parameters have values $\beta = 0$ [as a consequence of having
  assumed a two-integral DF $f(E, L_{z})$] and $\gamma = -0.32$,
  indicating a mild tangential anisotropy.
\end{enumerate}

These results are in good agreement with the analysis of {\galaxy} by
\citet{Koopmans2006}, which we here extend in a more rigorous way. In
particular, the results confirm the essentially isothermal profile of
the mass density distribution, which appears to be a defining
characteristic of early-type galaxies.

The best-fitting model is consistent with a dark matter fraction of
30~per cent within 10\,kpc (approximately corresponding to the
unprojected effective radius), similar to what \citet{Cappellari2006}
determine for the SAURON sample of early-type galaxies by making use
of three-integral Schwarzschild dynamical models under the caveat that
light traces mass. However, it seems plausible that a constant-M/L
model could still reproduce the observed kinematics of {\galaxy} if
one allowed for a more flexible three-integral dynamical model rather
than the two-integral model considered in this work. The relative
difficulty in unambiguously discriminating between the constant-$M/L$
model and the power-law model is also a consequence of the fact that
for the specific lens system studied here, the mass within the
Einstein radius is to a large extent dominated by the stellar mass, so
that the difference between the two models does not show up
dramatically in the lens model. The combined analysis of more distant
objects, however, is expected to provide more unequivocal results in
this respect. In general, for galaxies at higher redshift the ratio
between the Einstein radius and the effective radius becomes larger
(see e.g.~the SLACS galaxy sample in \citealt{Koopmans2006}) and any
deviation from a model for which mass follows light would become
significantly more prominent, since at least the surface brightness
slope would be much steeper than what is allowed by the lensing data.

The dynamical structure of {\galaxy} (i.e.\ anisotropy and
$\vphi/\bar{\sigma}$ map) is also in good agreement with what is found
by \citet{Cappellari2007} for the most massive ellipticals of the
SAURON sample.

The analysis has shown that the combination of gravitational lensing
and stellar dynamics is a powerful method which allows the dissection
in three dimensions of an elliptical galaxy (assumed to be well
described as a two-integral axisymmetric dynamical system), breaking
to a significant extent the classical degeneracies between inclination
and flattening as well as between mass and anisotropy. The way the
degeneracy between inclination and oblateness is overcome can be
understood within a simplified physical picture: $q$ and $i$ are
coupled in the projected potential which enters in the description of
gravitational lensing; when it comes to the dynamics, however, the
surface brightness weighted integral along the minor axis over the
line-of-sight stellar velocity dispersion is not a function of the
inclination, since, due to the properties of the two-integral DF, the
intersection of the velocity dispersion tensor with the $(v_{R},
v_{z})$ plane is always a circle, that is, $\sigma_{R}^{2} =
\sigma_{z}^{2}$ for each position in the galaxy (see
Section~\ref{sec:degeneracies}).

We have shown that the method, in its implementation as the {\dynlen}
algorithm, is robust enough to make use of observational data in order
to recover the non-linear parameters which characterize the total
gravitational potential and the geometry of the system (i.e.\
inclination, positional angle and lens centre) with relatively tight
error bars (the confidence intervals shown in
Table~\ref{tab:stat}). This first application therefore shows promise
for the future study of the other SLACS systems at higher redshift.

In forthcoming papers in this series we will extend this work to the
entire sample of 17~SLACS lenses with VLT VIMOS IFS, covering a range
of lens galaxy morphology, mass and redshift ($z = 0.08 - 0.35$). The
VLT sample will be complemented by a further 13 lenses for which we
have obtained long-slit spectroscopy at the Low Resolution Imager and
Spectrograph \citep[LRIS,][]{Oke1995} on the Keck-I telescope. Several
slit positions -- aligned with the major axis and offset along the
minor axis -- have been obtained for each system in the Keck sample,
thus effectively producing two-dimensional kinematic information
across most of the lens galaxy.

\section*{Acknowledgments}

We thank the anonymous referee for providing useful comments.  The
data published in this paper have been reduced using \textsc{vipgi},
designed by the VIRMOS Consortium and developed by INAF Milano. We
thank the developers of \textsc{vipgi}, in particular Paolo Franzetti,
for their support in using the software. We are grateful to Luca
Ciotti for enlightening discussion and to Giuseppe Bertin and James
Binney for valuable comments. We also thank Sebasti\'an S\'anchez for
technical assistance in preparing the observations.  MB acknowledges
the support from an NWO program subsidy (project number
614.000.417). OC and LVEK are supported (in part) through an NWO-VIDI
program subsidy (project number 639.042.505). We also acknowledge the
continuing support by the European Community's Sixth Framework Marie
Curie Research Training Network Programme, Contract
No. MRTN-CT-2004-505183 `ANGLES'.  TT acknowledges support from the
NSF through CAREER award NSF-0642621, by the Sloan Foundation through
a Sloan Research Fellowship, and by the Packard Foundation through a
Packard Fellowship.

\bibliography{SLACS-IFU-1}

\label{lastpage}

\clearpage

\end{document}